\documentclass[aps,10pt,physrev,onecolumn,groupedaddress,floatfix,longbibliography]{revtex4-2}

\usepackage{graphicx}
\usepackage[dvipsnames]{xcolor}
\usepackage{printlen}
\usepackage{amsmath,amssymb}
\usepackage{mathtools}
\usepackage{bbm}
\usepackage{natbib}
\usepackage{bm}
\usepackage{pifont}
\usepackage{booktabs}
\usepackage{array}
\usepackage[pdftex,breaklinks]{hyperref}
\hypersetup{
    colorlinks=true,
    citecolor=blue,
    linkcolor=blue,
    urlcolor=blue,
}
\usepackage{chemfig}
\allowdisplaybreaks

\begin{document}

\title{Solving stochastic gene expression models using queueing theory: a tutorial review}

\author{Juraj Szavits-Nossan}
\email{Juraj.Szavits.Nossan@ed.ac.uk}
\affiliation{School of Biological Sciences, University of Edinburgh, Edinburgh EH9 3JH, United Kingdom}

\author{Ramon Grima}
\email{Ramon.Grima@ed.ac.uk}
\affiliation{School of Biological Sciences, University of Edinburgh, Edinburgh EH9 3JH, United Kingdom}

\date{\today}

\begin{abstract}
Stochastic models of gene expression are typically formulated using the chemical master equation, which can be solved exactly or approximately using a repertoire of analytical methods. Here, we provide a tutorial review of an alternative approach based on queueing theory that has rarely been used in the literature of gene expression. We discuss the interpretation of six types of infinite server queues from the angle of stochastic single-cell biology and provide analytical expressions for the stationary and non-stationary distributions and/or moments of mRNA/protein numbers, and bounds on the Fano factor. This approach may enable the solution of complex models which have hitherto evaded analytical solution.  
\end{abstract}

\maketitle

\section*{Introduction}

Biochemical reaction systems are inherently stochastic, in the sense that even if we could account for all possible external factors, it is not possible to precisely predict which reaction event will occur in a small-time interval~\cite{gillespie1992rigorous}. The effect of this intrinsic noise on the dynamics of simple biochemical circuits has been extensively studied by means of the chemical master equation, a probabilistic description of reaction dynamics under the assumption that the time between successive reaction events is exponential (the Markovian memoryless assumption)~\cite{gillespie2007stochastic}. A repertoire of methods have been developed to approximately or exactly solve the chemical master equation in stationary and non-stationary conditions. These include the generating function method~\cite{mcquarrie1967stochastic,gardiner1985handbook,grima2012steady,shahrezaei2008analytical,veerman2018time}, the Poisson representation~\cite{gardiner1985handbook,gardiner1977poisson,iyer2014mixed,anderson2020time,Wang_2023}, the linear-noise approximation and its various extensions~\cite{van1992stochastic,grima2010effective,Thomas_2012,thomas2014phenotypic,Hufton_2016,Herath_2018} and methods inspired by quantum mechanics including ladder operators and Feynman-like diagrams~\cite{thomas2014system,vastola2021solving,harsh2023accurate}. These and other methods have been discussed in detail in various reviews~\cite{schnoerr2017approximation,weber2017master,bressloff2017stochastic,Gorin_2023}. In particular, due to the observed large cell-to-cell variation in mRNA and protein numbers~\cite{elowitz2002stochastic,zenklusen2008single}, there has been an intense interest in the application of these techniques to solve a large variety of models of transcription and/or translation~\cite{friedman2006linking,shahrezaei2008analytical,Ham_2020,Cao_2020,Peccoud_1995,kumar2014exact,grima2012steady,bokes2012exact,Zhou_2012,jia2022concentration,gorin2020special,thomas2014phenotypic,herbach2019stochastic}. The solution of each new stochastic model of gene expression is often laborious and specific to that model because the general form of the solution of the chemical master equation is only known for a class of chemical systems with rather restrictive constraints~\cite{van1976equilibrium,anderson2010product,cappelletti2016product}. The situation is made even more difficult by the fact that some gene expression systems require a non-Markovian description~\cite{Xu_2016,Jiang_2021,Fu_2022,kim2022systematic} for which very few analytical methods exist~\cite{leier2015delay}. Hence, there is considerable interest in the development of methods that circumvent the limitations of the present techniques.

A promising approach that has been put forward to solve complex reaction systems (biochemical or of another kind) is queueing theory. Queueing theory is a branch of mathematics which describes customers arriving to some facility where they receive service of some kind and then depart~\cite{Gross_2008}. The word ``queueing" describes a scenario in which there is a finite number of servers, so if all servers are busy, then new customers must wait or queue for the service. In biology, queueing theory has been used to solve models of enzymatic reactions~\cite{Levine_2007,Mather_2010,Mather_2011,Cookson_2011,Steiner_2016}, gene regulatory networks~\cite{Arazi_2004,Josic_2011,Dean_2022}, mRNA translation under limited resources~\cite{Mather_2013,Kulkarni_2013} and stochastic expression of a single gene~\cite{Elgart_2010,Jia_2011,Schwabe_2012, Kumar_2015,Choubey_2015,Thattai_2016,Horowitz_2017,Park_2018,Ali_2019,Kumar_2020,Shi_2020,Dean_2020,Yang_2022,Szavits_2023,Fralix_2023}. A variety of stochastic models of gene expression, in particular those describing expression occurring in bursts, have been mapped~\cite{Elgart_2010,Jia_2011,Kumar_2015} to a particular queueing system known as the $G^{X}/G/\infty$ queue~\cite{Liu_1990}, which was further reviewed in Ref.~\cite{bressloff2017stochastic}. Similarly, stochastic models of nascent RNA kinetics have been recently mapped to the $G/D/\infty$ queue~\cite{Szavits_2023} that is intimately connected to renewal theory~\cite{Cox_1967}. However, beyond these two queueing systems and few recent studies~\cite{Dean_2020,Dean_2022,Fralix_2023}, no further connection between gene expression modelling and queueing theory has been explored. 

In this review, we establish a deep connection between stochastic models of gene expression and queueing theory. We show that many stochastic models of gene expression can be easily solved using classical results from queueing theory, often without using the chemical master equation. Surprising, most of these results seem to be unknown to the gene expression modelling community. Therefore, we have structured this review as a tutorial so that that anyone, even without prior knowledge of queueing theory, can use it to analyse or solve their gene expression model of interest. Given the generality of this theory, it is clear that it can address the solution of much more complex models of gene expression than presently considered, and we hope this tutorial review gives readers the tools to achieve this goal. 

\section*{Queueing theory of stochastic gene expression}

\subsection*{A primer on queueing theory}

Before we establish the connection between stochastic models of gene expression and queueing theory, we briefly lay out the main characteristics that define a queueing system. These characteristics are: (1) the arrival process $A$, (2) the service process $S$, (3) the number of servers $c$, (4) the capacity of the queue $K$, (5) the calling population $N$, and (6) the queue's discipline $D$. These characteristics are usually summarized using the extended Kendall's notation $A/S/c/K/N/D$. 

The arrival process $A$ describes how often customers arrive at the system, and whether they arrive one at a time or in batches. Batch arrivals are typically denoted by the superscript $X$, as in $A^{X}$. The service process $S$ describes how long it takes to serve each customer. A given server may serve one customer at a time, or a batch of customers. The number of servers $c$ may be any number from $1$ to $\infty$. It is usually assumed that servers operate in parallel and are independent of each other. If all servers are busy, then new customers arriving at the system must queue for the service. The total number of customers in the system, which includes customers that are waiting and customers that are being served, is called the queue length. For infinite-server queues, the queue length is equal to the number of busy servers. The capacity $K$ of the queue is the maximum number of customers that are allowed to queue for the service at any given time. If the queue length reaches this number, then no further customers are allowed to join the queue until the queue length drops below this number due to service completion. The calling population $N$ is the total number of customers, which, if finite, may affect the arrival process. Finally, the queue's discipline $D$ describes how the next customer to be served is selected among the queueing population. A common example is the first come, first served discipline (FCFS), in which the customers are served in the order in which they arrive. The default values are $K=\infty$, $N=\infty$ and $D$ = FCFS, in which case a simpler notation $A/S/c$ is used instead of $A/S/c/K/N/D$.

Having laid out what constitute a queueing system, let us consider some examples. The simplest arrival process is a renewal process, in which the interarrival times are independent and identical random variables. Renewal processes are denoted by $G$ in Kendall's notation, which stands for general or unspecified interarrival time distribution. Special cases of renewal processes are the Poisson process denoted by $M$ (Markovian or memoryless), in which the interarrival times are exponential distributed, and the deterministic process denoted by $D$, in which the interarrival times are fixed. The simplest service process is one in which the service time of each customer is taken from the same probability distribution. A general or unspecified service time distribution is denoted by $G$, of which special cases are the exponential distribution denoted by $M$, and the deterministic (degenerate) distribution denoted by $D$. 

\subsection*{Stochastic gene expression as a queueing system}

Gene expression is a fundamental cellular process by which genetic information encoded by a gene is turned into a functional product, such as an RNA or protein molecule. Models of gene expression are typically concerned about the statistics of either RNA or protein counts as a proxy of gene activity; rarely the description of both is considered because the simultaneous measurement of RNA and protein in the same cell is challenging. In what follows, we will mostly focus on the RNA description of gene expression; where appropriate, we will discuss the protein description.

\subsubsection*{A general model for transcription and RNA degradation}

\begin{figure}[hbt]
    \centering
    \includegraphics[width=\textwidth]{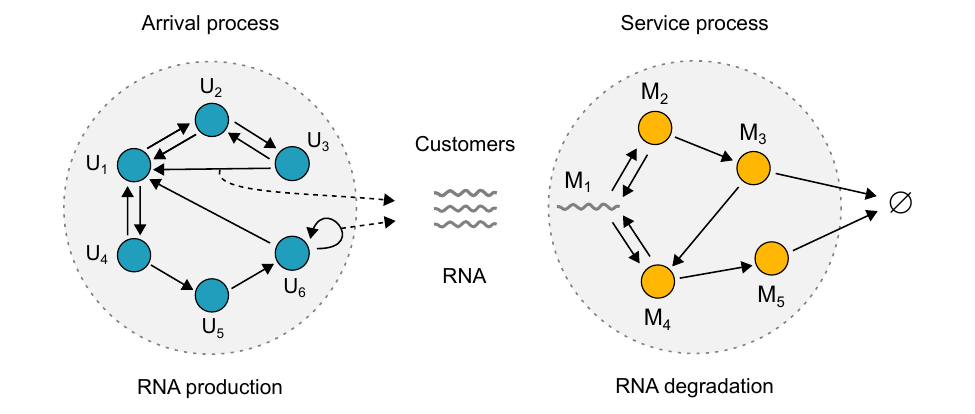}
    \caption{A general model of stochastic gene expression consisting of RNA production and degradation. The gene switches between multiple gene states, labelled by $U_1,\dots,U_S$, and eventually produces RNA, which is then processed and degraded through a process with multiple states labelled by $M_1,\dots,M_R$. Transitions between gene states are of two types, those that occur between two distinct gene states without the production of RNA (solid arrows without additional dashed arrows), and those that occur between two gene states, not necessarily distinct, and include the production of RNA (solid arrows with additional dashed arrows). The states and transitions presented here are for illustration purposes only and do not represent any particular RNA production or degradation mechanism.}    \label{fig1}
\end{figure}

Transcription---the synthesis of RNA---is typically modelled as a multistep process in which the gene switches between multiple states before it eventually produces an RNA molecule. Depending on the level of details, transitions between gene states may reflect individual biochemical events, such as binding of transcription factors and RNA polymerase at the promoter, or more phenomenologically, a combination of these events that results in the gene being either active or inactive. Once the RNA is produced, it goes through a series of steps until it is eventually degraded. In the Markovian setting, these steps can be described by the following reaction scheme
\begin{subequations}
\begin{alignat}{2}
    \label{transcription} & \text{(transcription):} && \quad U_{i}\xleftrightharpoons[k_{i,j}]{k_{j,i}}U_{j}, \quad U_{i}\xrightarrow{\rho_{i,j}}U_{j}+M_1,\quad i,j=1,\dots,S\\
    \label{degradation} & \text{(RNA degradation):} && \quad M_{i}\xleftrightharpoons[d_{i,j}]{d_{j,i}}M_{j},\quad M_i\xrightarrow{\lambda_i}\emptyset,\quad i,j=1,\dots,R,
\end{alignat}
\end{subequations}
where $U_1,\dots,U_S$ are gene states and $M_1,\dots,M_R$ are RNA species in various stages of the RNA degradation process (Fig.~\ref{fig1}). All transitions are assumed to be Markovian (memoryless) with constant (time-independent) rates. This is different from some models of gene expression in which degradation is mediated by enzymes~\cite{Parker_2004}, such that the effective degradation rate, obtained under timescale separation when the enzyme species are eliminated, is of the Hill form~\cite{Rao_2003, Thomas_2012}.

The reaction scheme in Eqs. (\ref{transcription}) and (\ref{degradation}) encompasses most of the gene expression models that have been proposed in the literature. These models are typically studied using the chemical master equation, which is an equation for the joint probability distribution $P_i(m_1,\dots,m_R,t)$ to find the gene in state $U_i$ at time $t$ with $m_1,\dots,m_R$ molecules of RNA species $M_1,\dots,M_R$, respectively. The master equation is rarely solved directly, instead it is turned it into an equation for the probability generating function $G_i(z_1,\dots,z_R)$ defined as
\begin{equation}
    G_i(z_1,\dots,z_R,t)=\sum_{m_1=0}^{\infty}\dots\sum_{m_R=0}^{\infty}P_i(m_1,\dots,m_R,t)z_{1}^{m_1}\dots z_{R}^{m_R}.
\end{equation}
Experimentally, what is measured is the total RNA number, irrespective of the degradation stage of RNA or the gene state. Hence, what we are ultimately interested in is $G(z,t)$ defined as
\begin{equation}
    G(z,t)=\lim_{z_1\rightarrow z}\dots\lim_{z_R\rightarrow z}\sum_{i=1}^{S}G_{i}(z_1,\dots,z_R,t)\equiv \sum_{m=0}^{\infty}P(m,t)z^m,
\end{equation}
where $P(m,t)$ is the probability distribution of the total RNA number $m=m_1+\dots+m_R$ at time $t$. In the chemical master equation formalism, we first set up the chemical master equation for $P_i(m_1,\dots,m_R,t)$, from which we derive the equation for the probability-generating function $G_i(z_1,\dots,z_R,t$. We then solve the equation for $G_i$ for each $i=1,\dots,S$, after which we add all $G_i(z_1,\dots,z_R,t)$ and set $z_1=\dots=z_R$ to get $G(z,t)$. Generally, this is a tedious procedure that needs to be repeated for each new model. Later we will show how in some cases queueing theory can bypass these difficulties and give the queue length distribution directly without using the chemical master equation formalism.

We now reformulate the reaction scheme in Eqs. (\ref{transcription}) and (\ref{degradation}) in terms of queueing theory. We consider the ``customers" to be RNA molecules that arrive according to Eq. (\ref{transcription}), and are serviced according to Eq. (\ref{degradation}). We do not count transient RNA molecules as separate RNA species. Instead, we consider $M_1,\dots,M_R$ to be transient states of the same RNA molecule. This is equivalent to saying that the queue length is equal to the total number of RNA molecules, irrespective of the stage of their service. According to Eq. (\ref{degradation}), once an RNA molecule is produced, it is immediately available to degradation machinery. Since we assume the degradation process to be the same for each RNA molecule, the number of servers $c$ is infinite. The capacity of the queue $K$ and the calling population $N$ are also assumed to be infinite. The queueing discipline $D$ does not apply to an infinite-server queue, as there is no queue, only the number of busy servers. However, since RNAs begin their service in the order in which they are produced, we set $D$ = FCFS. This, of course, does not mean the RNA that was produced first will degrade first, unless the degradation process is deterministic taking a fixed amount of time. Based on these characteristics, we conclude that stochastic gene expression is equivalent to a queueing system $A/S/\infty$, where $A$ is the arrival process, $S$ is the service process, and there are infinitely many servers (since $K$, $N$ and $D$ take their default values, they are omitted from the notation).

\subsubsection*{Transcription as an arrival process}

We now look more closely into Eq. (\ref{transcription}) describing transcription. Eq. (\ref{transcription}) describes what is known as a Markovian arrival process (MAP)~\cite{Neuts_1979}. The MAPs are important for modelling arrivals because their set is dense in the set of all the stationary point processes~\cite{Asmussen_1993}, which means that any stochastic process consisting of discrete events occurring at random times can be well approximated by a MAP. Another quality of the MAPs is their Markovian nature, which makes them mathematically tractable. The dynamics of the MAP described by Eq. (\ref{transcription}) can be expressed using two $S\times S$ matrices $D_0$ and $D_1$ defined as
\begin{equation}
    \label{MAP-D0-D1}
    [D_0]_{i,j}=\begin{dcases}
        -\sum_{\genfrac{}{}{0pt}{}{j=1}{j\neq i}}^{S}k_{i,j}-\sum_{j=1}^{S}\rho_{i,j}, & i=j\\ 
        k_{i,j}, & i\neq j\end{dcases},\qquad [D_1]_{ij}=\rho_{i,j}.
\end{equation}
The matrix $D_0$ accounts for transitions during which no RNA is produced, whereas the matrix $D_1$ accounts for transitions that result in the production of RNA. Note that $D_0+D_1$ is the transition matrix of the process that tracks only gene states, but not the production of RNA. In case of batch arrivals, there are additional matrices $D_k$ for each batch size $k\geq 1$, where $[D_k]_{i,j}=a_k \rho_{i,j}$ and $a_k$ is the batch size distribution (the probability that the batch is of size $k$) such that $\sum_{k=1}^{\infty}a_k=1$. The MAP with batch arrivals is called the batch Markovian arrival process (BMAP).

We mention two important properties of the MAP: (1) that its interarrival times are phase-type distributed, and (2) that its successive interarrival times are generally correlated. The first property means that any interarrival time distribution can be approximated by a MAP, since the set of phase-type distributions is dense in the set of all continuous distributions of non-negative random variables~\cite{Asmussen_2000}. The second property implies that there is a memory between successive RNA production events. To understand where this memory comes from, we say that a gene state $U_i$ is an active state if an RNA molecule can be produced from it, i.e. if $\rho_{i,j}\neq 0$, where $j\in\{1,\dots,S\}$. For a given active state $U_i$, let us denote by $\kappa_{i,j}=\rho_{i,j}/\sum_{j=1}^{S}\rho_{i,j}$ the probability that the gene switches from $U_i$ to $U_j$ after producing an RNA molecule. If the gene produced an RNA molecule from state $U_i$ at time $0$, then the probability density function of the time $t$ until the next RNA is produced is given by
\begin{equation}
    \label{MAP-interarrival-pdf-i}
    f_{i}(t)=\sum_{j=1}^{S}\sum_{k=1}^{S}\sum_{l=l}^{S}\kappa_{i,j}\left[e^{D_0 t}\right]_{j,k}W_1(k\rightarrow l)=\bm{\kappa}_i e^{D_0 t}D_1 \bm{1}^T=\bm{\kappa}_i e^{D_0 t}(-D_0 \bm{1}^T).
\end{equation}
which is a phase-type distribution described by the initial vector $\bm{\kappa}_i=(\kappa_{i,1},\dots,\kappa_{i,S})$ and the transition matrix $D_0$. The time until the next RNA molecule is produced thus depends on the gene state the previous RNA molecule was produced from. Similarly, the joint probability density function of $t_n$ and $t_{n+1}$ reads $f_{n,n+1}(t_n,t_{n+1})=\bm{\pi}_{n-1}e^{D_0 t_{n}}D_1e^{D_0 t_{n+1}}D_1\bm{1}^T$, i.e. the successive interarrival times are mutually correlated. On the other hand, if $\bm{\kappa}_i\equiv\bm{\kappa}$ is independent of $i$, meaning that the gene resets according to the same probability vector $\bm{\kappa}$ after each RNA production event, then the MAP becomes a renewal process $G$. In that case, the successive interarrival times are independent and identically distributed random variables whose  probability density function is 
\begin{equation}
    \label{G-interarrival-pdf}
    f(t)=\bm{\kappa} e^{D_0 t}D_1 \bm{1}^T=\bm{\kappa} e^{D_0 t}(-D_0 \bm{1}^T),
\end{equation}
Since the $i$th row of $D_1$ is equal to $\bm{\kappa}_i$ multiplied by $\sum_j\rho_{i,j}=[D_0\bm{1}^T]_{i,1}$, then $\bm{\kappa}_i$ being independent of $i$ implies that 
\begin{equation}
    \label{renewal-condition}
    D_1=(-D_0\bm{1}^T)\bm{\kappa}.
\end{equation}
Hence, if $D_1$ is made up of rows that are all equal up to a scaling factor, then the two successive interarrival times in the MAP are uncorrelated, and the MAP becomes a renewal process. We will refer to $\bm{\kappa}_i$ being independent of $i$ and its consequence in Eq. (\ref{renewal-condition}) as the renewal condition of the MAP. The sufficient condition for the MAP to be a renewal process is if there is only one active state~\cite{Fischer_1993}. 

\begin{figure}[hbt]
    \centering
    \includegraphics[width=\textwidth]{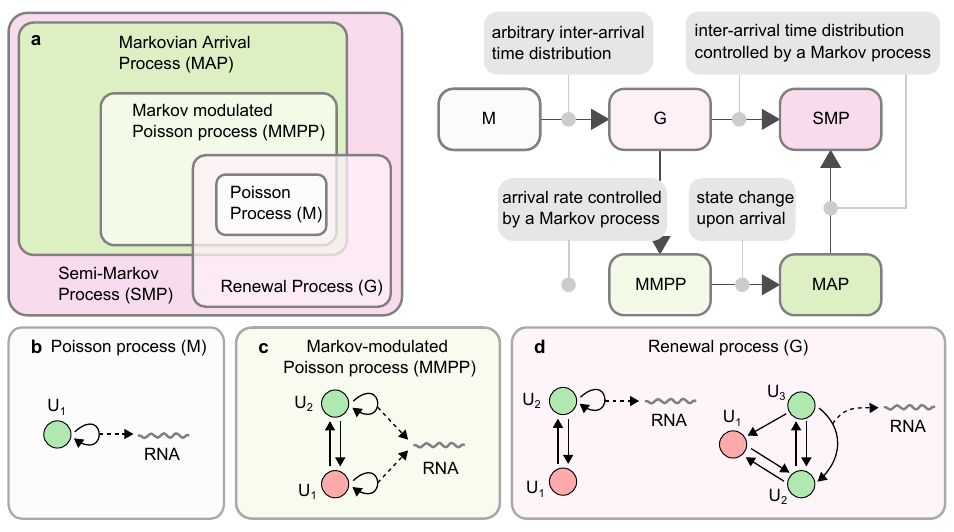}
    \caption{Arrival processes related to the Markovian arrival process (MAP) and the equivalent models of RNA production. (a) Left: Venn diagram showing relationships between different arrival processes. Right: different ways of generalizing the Poisson process to include non-exponential and correlated interarrival times. The simplest MAP is the Poisson process ($M$). A Poisson process whose rate is controlled by a Markov process is called the Markov-modulated Poisson process (MMPP). The MAP is a generalization of the MMPP to allow state change upon arrival. A generalization of the Poisson process to an arbitrary interarrival time distribution while keeping interarrival times uncorrelated is the renewal process ($G$). A renewal process whose interarrival time distribution is controlled by a Markov process is called the semi-Markov process (SMP). The MAP is a special case of the SMP, hence the SMP is more general. (b)-(d) Examples of Markovian arrival processes in the context of gene expression. (b) The Poisson process, describing a gene that is always active. (c) The leaky telegraph process, describing a gene that switches between two states and produces RNA from both states~\cite{Kepler_2001}. The leaky state (denoted by red) is typically much less active. (d) Left: the telegraph process, describing a gene that switches between two states of activity and inactivity. Right: a three-state process that accounts for RNA polymerase recruitment and its release into productive elongation~\cite{Bartman_2019, Braichenko_2021}.}
    \label{fig2}
\end{figure}

Fig.~\ref{fig2} summarizes various stochastic processes that are related to the MAP. The simplest MAP is the Poisson process (denoted by $M$ for Markovian or memoryless), which has only one state. This process describes a gene that is always active and produces RNA at exponentially distributed intervals [Fig.~\ref{fig2}(b)]. One way to generalize the Poisson process is to have the arrival rate controlled by a finite-state Markov process. This process, which is called the Markov modulated Poisson process (MMPP), is a special case of the MAP in which $D_1$ is a diagonal matrix~\cite{Fischer_1993}. The simplest stochastic gene expression model with this arrival process is the leaky telegraph model~\cite{Kepler_2001}, in which the gene switches between two states, both of which are transcriptionally active [Fig.~\ref{fig2}(c)]. We note that in the MMPP, the gene remains in the same state immediately after producing RNA. A gene that produces RNA from multiple states, but is allowed to change state upon the production of RNA (in which case $D_1$ is no longer a diagonal matrix), is described by a general MAP.   

Another way to generalize the Poisson process is to allow for non-exponential interarrival times, while keeping the interarrival times uncorrelated. This defines a renewal process (denoted by $G$ for general or arbitrary interarrival distribution, or $GI$ to emphasize that the interarrival times are mutually independent). An example of stochastic gene expression model with this arrival process is the (random) telegraph model [Fig.~\ref{fig2}(d)]. Another example is a three-state model that accounts for the binding of RNA polymerase and its release into productive elongation~\cite{Bartman_2019,Braichenko_2021}. Here gene state changes upon arrival since the released RNA polymerase is lost and a new RNA polymerase needs to be recruited at the promoter. As we have shown above, the MAP is not a renewal process, unless $D_1$ takes a special form stated in Eq. (\ref{renewal-condition}). This condition requires that the gene resets to the same initial probability vector (of being in a given gene state) after each RNA production event. Any MAP with a single active state is therefore a renewal process. A sufficient (but not necessary) condition for an MMPP to be a renewal process is that all but one diagonal elements of $D_1$ are zero (or that the rank of $D_1$ is 1), which means that the gene always produces RNA from the same state. This immediately implies that gene expression models with multiple active states, such as the leaky telegraph model, predict correlated interarrival times, i.e. transcriptional memory. Such memory is absent in models with a single active state. 

Finally, we mention a generalization of the renewal process in which the interarrival time distribution itself is controlled by a Markov process, which is called the semi-Markov process (SMP). A semi-Markov process is defined as a sequence of random variables $(X_n, T_n)$, where $T_n$ is time of the $n$-th arrival, and $X_n$ is the state of the system in the time interval $[T_n, T_{n+1}\rangle$. Given $X_n=i$, the interarrival time $t_{n+1}=T_{n+1}-T_n$ and the new state $X_{n+1}=j$ are selected according to the probability $P(t_{n+1}\leq t, X_{n+1}=j\vert X_n=i)=Q_{ij}(t)$. The MAP is a special case of the SMP with the following conditional probability matrix $Q_{ij}(t)$~\cite{Fischer_1993},
\begin{equation}
    \label{MAP-to-SMP}
    Q_{ij}(t)=\left[\int_{0}^{t}dt'e^{D_0 t'}D_1\right]_{ij}=\left[\left(I-e^{D_0 t}\right)(-D_{0}^{-1}D_1).\right]_{ij}.
\end{equation}
When mapping the MAP to the SMP, only the states at the arrival epochs are recorded. These states form what is known as the embedded Markov chain, whose probability transition matrix is $D_{0}^{-1}D_1$. Depending on the matrices $D_0$ and $D_1$, some states of the MAP may appear as transient states of the embedded Markov chain. These states do not appear at the arrival epochs, but they do leave an imprint in the interarrival time distributions of the SMP. Therefore, the MAP is preferred over the SMP when we want to give the process between arrivals a Markovian interpretation, whereas the SMP is preferred over the MAP when we have limited information about the process between arrivals.

\subsubsection*{RNA degradation as a service process}

The RNA degradation process described by Eq. (\ref{degradation}) is a Markov process that consists of $R$ transient states ($M_1,\dots,M_R$) and one absorbing state ($\emptyset$). The transition matrix of this process is given by
\begin{equation}
    [D_{\text{deg}}]_{i,j}=\begin{dcases}-\lambda_i-\sum_{\genfrac{}{}{0pt}{}{j=1}{j\neq i}}^{R}d_{i,j}, & i=j\\
    d_{i,j}, & i\neq j\end{dcases},
\end{equation}
The probability density function of the time it takes to degrade an RNA molecule starting from state $M_1$ is given by
\begin{equation}
    \label{RNA-degradation-pdf}
    h(t)=\bm{e}_1 e^{D_{\text{deg}}t}(-D_{\text{deg}}\bm{1}^T),
\end{equation}
where $\bm{e}_1=(1,0,\dots,0)$ is the initial probability vector (RNA degradation always starts in state $M_1$), $\bm{1}=(1,\dots,1)$ and $\bm{1}^T$ is the transpose of $\bm{1}$. The distribution in Eq. (\ref{RNA-degradation-pdf}) is known as a phase-type distribution of order $R$ and is denoted by $PH$ in queueing theory. The appeal of phase-type distributions is that their set is dense in the field of all positive-valued distributions, meaning that the distribution of any positive-valued random variable can be well approximated by a phase-type distribution~\cite{Asmussen_2000}. This, and their mathematical tractability, makes phase-type distributions capable of modelling complex service patterns. Well-known examples of phase-type distributions include the exponential distribution (denoted by $M$), $h(t)=\lambda e^{-\lambda t}$, and the Erlang distribution with shape $R$ (denoted by $E_R$), $h(t)=\lambda^R t^{R-1}e^{-\lambda t}/(R-1)!$, which is the distribution of a sum of $R$ identical exponentially distributed random variables. Many models of gene expression assume an exponential distribution of the RNA degradation times, which is equivalent to assuming a single rate-limiting step. In contrast, the Erlang distribution models a service process where there are many fast steps but in which only $R$ steps are rate-limiting. We note that the mean and the variance of the Erlang distribution are $R/\lambda$ and $R/\lambda^2$, respectively, which gives the coefficient of variation $CV=1/\sqrt{R}$. In fact, the Erlang distribution has the smallest coefficient of variation among all phase-type distributions of the same order, which makes it most suitable for modelling deterministic service times. Indeed, if we fix the mean $T=R/\lambda$, and set $R\rightarrow\infty$ and $\lambda\rightarrow\infty$, we get the deterministic distribution, $h(t)=\delta(t-T)$, which is denoted by $D$ in queueing theory. Hence, we can use the deterministic distribution to approximate a service process that consists of many similar fast steps. One such example is transcription elongation, during which RNA polymerase traverses thousands of nucleotides and produces nascent RNA, one nucleotide at a time. In this case, the customers are nascent RNAs, the arrival process is transcription initiation, and the service process are the processes of transcriptional elongation and termination~\cite{Xu_2016, Szavits_2023}.

\begin{figure}[htb]
    \centering
    \includegraphics[width=\textwidth]{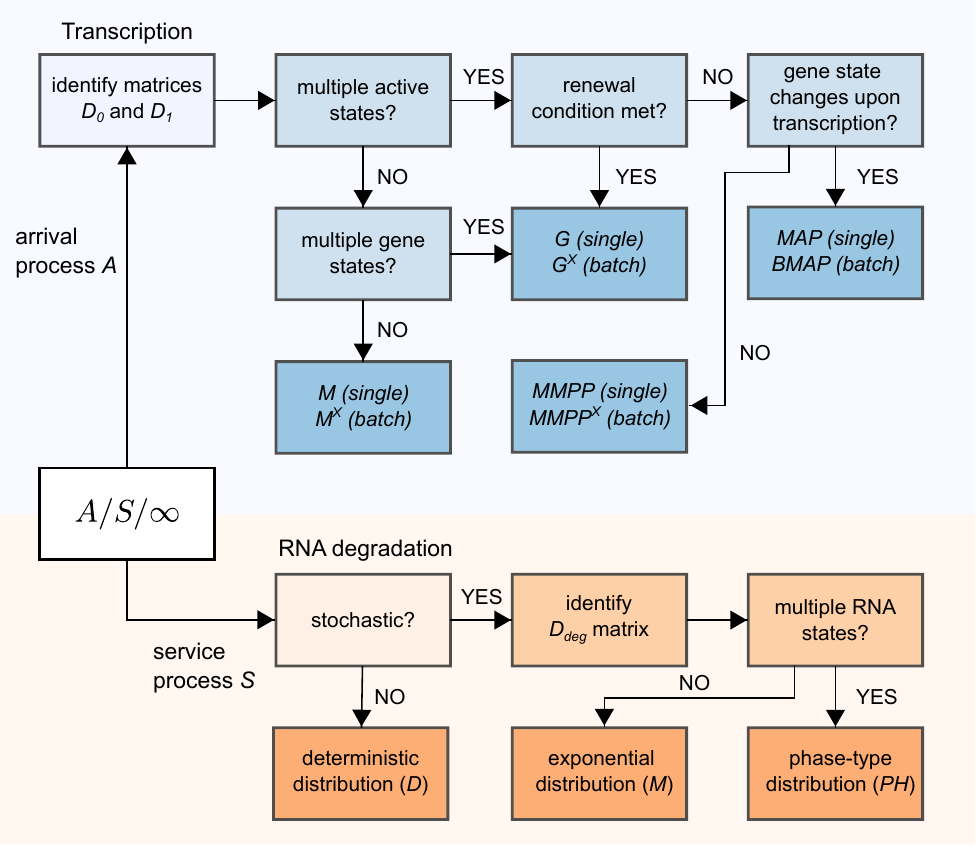}
    \caption{A flow chart for identifying the arrival process describing transcription (top) and the service process describing RNA degradation (bottom) for a given realization of the general model described by Eqs. (\ref{transcription}) and (\ref{degradation}).}
    \label{fig3}
\end{figure}

\section*{The landscape of infinite-server queues}

In the previous section, we considered a multistep model of RNA production and degradation, and showed that it can be mapped to an infinite-server queue $A/S/\infty$, where transcription is the arrival process $A$, RNA degradation is the service process $S$, and the number of observed RNA is the queue length (the number of busy servers). We showed that the arrival process is described by the Markovian arrival process ($MAP$), which includes the Poisson process (denoted by $M$) and the Markov-modulated Poisson process (denoted by $MMPP$) as special cases. We derived the renewal condition under which the MAP becomes a renewal process (denoted by $G$). Finally, we showed that RNA degradation is fully specified by the RNA degradation time distribution, which is assumed to be the same for all RNA. We showed that the distribution of the RNA degradation time is a phase-type distribution ($PH$), of which an exponential distribution ($M$) is a special case, and that a deterministic (degenerate) distribution ($D$) is a good approximation for service times that include numerous rate-limiting steps. In Fig.~\ref{fig3}, we present a flow diagram which can be used to identify the arrival and service processes, once the matrices $D_0$, $D_1$ and $D_{\text{deg}}$ describing the model in Eqs. (\ref{transcription}) and (\ref{degradation}) are identified. 

\begin{table}[htb]
    \centering
    \caption{A summary of known results for selected infinite-server queues that are relevant for stochastic gene expression modelling. The table refers to the non-stationary and stationary RNA number distributions and their corresponding moments.}
    \begin{tabular}{
        m{0.12\textwidth}
        >{\centering}m{0.09\textwidth}
        >{\centering}m{0.09\textwidth}
        >{\centering}m{0.09\textwidth}
        >{\centering}m{0.09\textwidth}
        >{\centering}m{0.14\textwidth}
        >{\centering\arraybackslash}m{0.19\textwidth}}
        \toprule
        & \multicolumn{4}{c}{Renewal arrivals} & \multicolumn{2}{c}{Non-renewal arrivals}\\
        \cmidrule(r){2-5}\cmidrule(l){6-7}
        Queue & $G^X/G/\infty$ & $G/M/\infty$ & $M^X/G/\infty$ & $G/D/\infty$ & $MMPP/M/\infty$ & $MMPP/G/\infty$\\
        Reference &~\cite{Liu_1990} &~\cite{Takacs_1958} &~\cite{Liu_1990} &~\cite{Cox_1967,Liu_1990} &~\cite{Ocinneide_1986} &~\cite{Blom_2014}\\
        \midrule
        Moments (ns) & \textcolor{Green}{\ding{51}} (RR) & \textcolor{Green}{\ding{51}} (LT) & \textcolor{Green}{\ding{51}} & \textcolor{Green}{\ding{51}} (LT) & \textcolor{Green}{\ding{51}} (RR) & \textcolor{Green}{\ding{51}} (mean and variance)\\
        Prob. dist. (ns) & \textcolor{Red}{\ding{55}} (TS) & \textcolor{Green}{\ding{51}} (LT) & \textcolor{Green}{\ding{51}} & \textcolor{Green}{\ding{51}} (LT) & \textcolor{Red}{\ding{55}} (TS) & \textcolor{Red}{\ding{55}}\\
        Moments (s) & \textcolor{Green}{\ding{51}} (RR) & \textcolor{Green}{\ding{51}} \hphantom{(LT)} & \textcolor{Green}{\ding{51}} & \textcolor{Green}{\ding{51}} (LT) & \textcolor{Green}{\ding{51}} (RR) & \textcolor{Green}{\ding{51}} (mean and variance)\\
        Prob. dist. (s) & \textcolor{Red}{\ding{55}} (TS) & \textcolor{Green}{\ding{51}} \hphantom{(LT)} & \textcolor{Green}{\ding{51}} & \textcolor{Green}{\ding{51}} (LT) & \textcolor{Red}{\ding{55}} (TS) & \textcolor{Red}{\ding{55}} \\
        \cmidrule{1-7}
        \multicolumn{7}{l}{Symbols: ns = non-stationary, s = stationary, LT = Laplace transform, RR = recurrence relation, TS = truncated series}\\
        \bottomrule
    \end{tabular}
    \label{tab1}
\end{table}

In this section, we review known results for six infinite-server queues made by combining the arrival and service processes described above, which are of particular importance for stochastic gene expression modelling. We focus on queues whose arrivals are described by renewal ($G$) and Markov-modulated processes ($MMPP$), as these type of arrivals are present in most of the stochastic gene expression models in the literature. The main results are summarized in Table~\ref{tab1}. For each queueing system, we report whether the non-stationary and stationary queue length distributions and their corresponding moments are known, along with a reference where these results can be found. Some results are in a closed form, whereas others require inverting the Laplace transform (denoted by LT), computing the moments by recursive relations (denoted by RR) or approximating the probability distribution by truncated series (TS). In the subsection below, we discuss these results in detail for the $G^X/G/\infty$, $G/M/\infty$, $M^X/G/\infty$, $G/D/\infty$ and $MMPP/M/\infty$ queues. We do not show results for the $MMPP/G/\infty$ queue, as they are quite complicated, and only the mean and the variance have been reported. Other infinite-server queues not mentioned in Table~\ref{tab1} are discussed later. 

The advantage of queueing theory over the traditional chemical master equation will become particularly clear for queues with renewal arrivals (the first four queues in Table~\ref{tab1}). In those queues, the information about the arrival process is fully encoded in the interarrival time distribution, which can be easily computed from the chemical master equation. The results for these queues are expressed in terms of general (arbitrary) interarrival time distribution, hence solving the full chemical master equation for the joint distribution of molecule numbers is not necessary. This in turn allows one not only to get analytical results more quickly and elegantly for large classes of gene expression models, but in some cases to get general results that would otherwise be difficult to infer from solving individual models one by one using the chemical master equation. In the following subsections, we have compiled the most useful of these results, whereas further details can be found in the accompanying Appendices.

\subsection*{\texorpdfstring{$G^X/G/\infty$}{G\^X/M/infinity} queue}

The $G^X/G/\infty$ queue is an infinite-server queue in which the arrivals constitute a renewal process, interarrival times are independent and identically distributed random variables with a general (arbitrary) distribution, customers arrive in batches with a general (arbitrary) batch-size distribution, and the service times have a general (arbitrary) distribution. The model of gene expression described by Eqs.~(\ref{transcription}) and (\ref{degradation}) is a special case of the $G^X/G/\infty$ queue, provided $D_0$ and $D_1$ satisfy the renewal condition in Eq.~(\ref{renewal-condition}), and RNAs are produced one by one (i.e. in batches of fixed size $1$). This model accounts for both multistep transcription and multistep RNA degradation, and as such includes many models of gene expression, most of which have simple one-step RNA degradation. Examples include the telegraph model~\cite{Peccoud_1995}, various three-state models~\cite{Suter_2011,Bartman_2019,Cao_2020}, and the ratchet model~\cite{Schwabe_2012,Zhou_2012}.

The $G^X/G/\infty$ has been studied in Ref.~\cite{Liu_1990} as a generalization of the $G/G/\infty$ queue~\cite{Takacs_1958} to batch arrivals. When both interarrival and service time distributions are arbitrary, the moments of the queue length distribution must be computed recursively, starting from the first moment. The queue length distribution, which can be expressed as a series involving binomial moments, is generally not known, unless all the binomial moments can be computed (see Appendix~\ref{appendix-a} for details). Special cases where the moments can be computed explicitly are the $G/M/\infty$ and $M^X/G/\infty$ queues, which we discuss separately. Another special case is $G/D/\infty$ queue ($D$ is for deterministic service), for which the moments can be computed by inverting Laplace transform with respect to the fixed service time.

It is interesting to note that the $G^X/G/\infty$ queue and Ref.~\cite{Liu_1990} have been the sole point of reference for most of the literature connecting stochastic gene expression to queueing theory~\cite{Jia_2011,Kumar_2015,bressloff2017stochastic}.That is in our opinion unfortunate, because the results for the $G^X/G/\infty$ queue (in its general setting) are limited to the moments of the queue length distribution, whereas the queue length distribution itself remains elusive. This explains why queueing theory has so far played a minor role in analysing stochastic models of gene expression. On the other hand, if one sacrifices the generality of the $G^X/G/\infty$ queue, and instead considers its special cases---$G/M/\infty$, $M^X/G/\infty$ and $G/D/\infty$ queues, all of which are undoubtedly relevant for gene expression modelling---then for those queues it is possible to compute both non-stationary and stationary queue length distributions \textit{without} using the chemical master equation. To the best of our knowledge, this fact has been largely overlooked in the biological modelling community.

\subsection*{\texorpdfstring{$G/M/\infty$}{G/M/infinity} queue}

\begin{figure}[hbt]
    \centering
    \includegraphics[width=\textwidth]{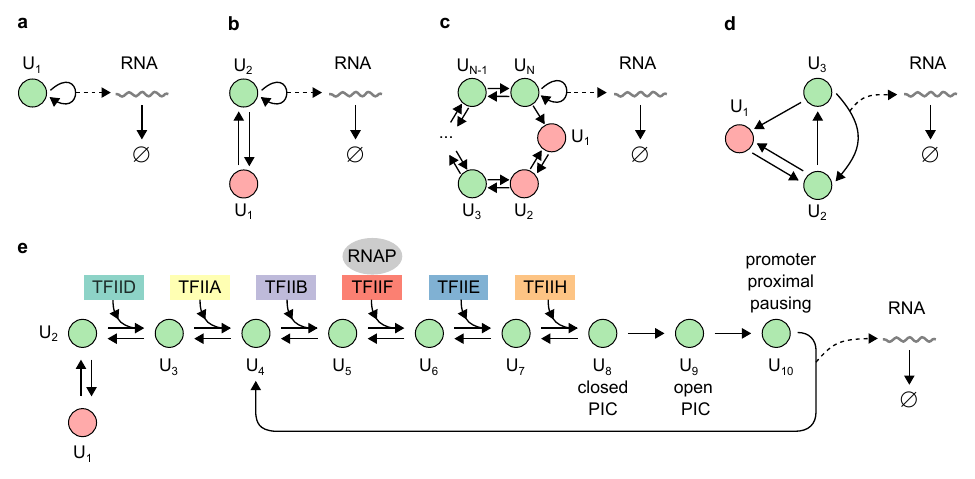}
    \caption{Examples of stochastic gene expression models that are equivalent to the $G/M/\infty$ queue, in the order of increased complexity. The red states represent inactive states, whereas the green states represent states belonging to the transcription initiation pathway. (a) The birth-death model, describing a gene that is always on. (b) The telegraph model, describing a gene that switches between two states of activity and inactivity. (c) The ratchet model, which is a generalization of the telegraph model to more than two states, of which only one is transcriptionally active~\cite{Schwabe_2012,Zhou_2012}. (d) A three-state model that accounts for the binding of RNA polymerase and its release into productive elongation after unpausing from the proximal promoter paused state~\cite{Bartman_2019, Cao_2020}. (e) A canonical model of transcription initiation~\cite{Kingston_1994,Sainsbury_2015} that accounts for the on and off switching of the promoter, the binding and unbinding of six general transcription factors (IID, IIA, IIB, IIF, IIE and IIH) and RNA polymerase (RNAP), the unwinding of the double-stranded DNA by the preinitiation complex (PIC), and the promoter proximal pausing of RNA polymerase in metazoans~\cite{Adelman_2012,Jonkers_2014}. After initiation, the gene returns to an earlier state in the transcription initiation pathway as it loses some of its transcription factors~\cite{Zawel_1995,Yean_1999}. In all five examples, the renewal condition is met in a deterministic fashion: after an RNA production event, the same state is always chosen.}
    \label{fig4}
\end{figure}

The $G/M/\infty$ queue is an infinite-server queue in which the interarrival times are independent and identically distributed random variables, customers arrive individually one by one, and the service times are exponentially distributed. It is a special case of the $G^X/G/\infty$ queue with batches of size $1$ and exponential service times. Many models of gene expression can be mapped to this queue, some of which are shown in Fig.~\ref{fig4}. The simplest is the one-state (birth-death) process in which the gene is always active and produces RNA at exponential intervals [Fig.~\ref{fig4}(a)]. The popular telegraph model in which the gene switches between two states of activity and inactivity and produces RNA from the active state is shown in Fig.~\ref{fig4}(b)~\cite{Peccoud_1995}. Fig.~\ref{fig4}(c) shows the ratchet model, which is a generalization of the telegraph model to multiple transcriptionally inactive states that are accessed sequentially~\cite{Schwabe_2012,Zhou_2012}. These three models have in common that the gene remains in the active state upon the production of RNA. In contrast, Fig.~\ref{fig4}(d) shows the refractory model which accounts for the binding of RNA polymerase and its release into productive elongation, after which the gene switches back to an earlier state absent of RNA polymerase~\cite{Bartman_2019,Cao_2020}. Finally, Fig.~\ref{fig4}(e) shows a canonical model of eukaryotic transcription~\cite{Kingston_1994,Sainsbury_2015,Szavits_2023} that includes the on and off switching of the promoter, the binding of six general transcription factors (IID, IIA, IIB, IIF, IIE and IIH) and RNA polymerase, the unwinding of the double-stranded DNA, and the promoter proximal pausing of RNA polymerase in metazoans~\cite{Adelman_2012,Jonkers_2014}.

A unique property of $G/M/\infty$ queue is that its stationary queue length distribution depends only on the Laplace transform of the interarrival time probability density ($\phi(s)$) and the service rate ($\lambda$). Hence, once the Laplace transform $\phi(s)$ is computed, the stationary queue length distribution and all of its moments follow immediately. The same is true for the non-stationary case, except that the calculations are more involving and include the inverse Laplace transform. These results were first derived in Ref.~\cite{Takacs_1958}, and are summarized in Appendix~\ref{appendix-b}. For example, the Fano factor (the ratio of the variance and the mean) reads
\begin{equation}
    \label{FF-GMI}
    FF=\frac{1}{1-\phi(\lambda)}-\frac{1}{\alpha\lambda},
\end{equation}
where $\alpha$ is the mean interarrival time, $\alpha=-\phi'(0)$. If we apply general bounds on the Laplace transform $\phi(s)$~\cite{Eckberg_1977}, we get that the Fano factor is bounded between $1/2$ and $1+CV_{a}^{2}$, where $CV_{a}$ is the coefficient of variation of the interarrival time distribution (the ratio of standard deviation and mean),
\begin{equation}
    \frac{1}{2}\leq FF\leq 1+CV_{a}^{2}.
\end{equation}
Hence, when arrivals are renewal and service is exponential, the Fano factor cannot be smaller than $1/2$, regardless of the arrival process and the service rate. For example, several cell division genes in fission yeast that have been recently reported exhibiting sub-Poissionian fluctuations of mRNA numbers all have the Fano factor above the lower bound of $1/2$, and therefore can be described by the $G/M/\infty$ queue \cite{weidemann2023minimal}. 

We now apply the $G/M/\infty$ queue to the gene expression model described by Eqs.~(\ref{transcription}) and (\ref{degradation}), under the renewal condition in Eq.~(\ref{renewal-condition}). In that case, the interarrival times are phase-type distributed, and their probability density function , $f(t)$, is given by Eq.~(\ref{G-interarrival-pdf}). Alternatively, $f(t)$ can be computed from the chemical master equation in which the RNA production event is treated as a transition into an absorbing state, after which the process resets itself. The Laplace transform of $f(t)$ in Eq.~(\ref{G-interarrival-pdf}) is given by Eq.~(\ref{LT-f-GMI}) in Appendix~\ref{appendix-b}. Using the fact that $\phi(s)$ is a rational function of $s$, it can be shown that the probability generating function of the stationary queue length distribution $P(m)$ can be written as,
\begin{equation}
    \label{G-PHMI}
    G(z)={}_p F_{q}\left(\frac{a_1}{\lambda},\dots,\frac{a_p}{\lambda};\frac{b_1}{\lambda},\dots,\frac{b_q}{\lambda};c \lambda^{p-q-1}\left(z-1\right)\right),
\end{equation}
where $c$, $p$, $q$, $a_1,\dots,a_p$, and $b_1,\dots,b_q$ are implicitly defined in Eq.~(\ref{LT-f-pars}). From here, it is straightforward to compute the stationary queue length distribution $P(m)$ and its moments using properties of the generalized hypergeometric function ${}_p F_{q}(a_1,\dots,a_p;b_1,\dots,b_q;z)$, see Eq.~(\ref{Pm-PHMI}). For example, this expression can be used to obtain the stationary RNA number distributions for the models in Fig.~\ref{fig3}(a), (b), (c) and (d), which have been previously derived using the master equation approach~\cite{Peccoud_1995, Zhou_2012, Braichenko_2021}. In Appendix~\ref{appendix-b}, we show this calculation for the telegraph model. 

Another general result that applies to models of gene expression with phase-type renewal arrivals and exponential service concerns the transient behaviour of the mean RNA number. If we assume that no RNA is present at time $t=0$, then for short times the mean RNA number follows a power law,
\begin{equation}
    \label{general-m-t}
    \langle m(t)\rangle=\frac{A}{n!}t^{n}+O(t^{n+1}),
\end{equation}
where the exponent $n$ is equal to the minimal number of gene states that are visited from the initial gene state at time $t=0$ until the synthesis of the first mRNA molecule~\cite{Nicoll_2023}. An experiment in which this type of initial condition is typically met is gene induction, in which a gene is initially inactive and is subsequently activated. Hence, measuring $\langle m(t)\rangle$ after induction can help us to estimate the number of regulatory steps in transcription. Recent applications of this result to experimental data in yeast and mouse are consistent with gene expression models that have multiple inactive gene states rather than a single inactive state assumed by the telegraph model~\cite{Nicoll_2023}.

\begin{figure}[hbt]
    \centering
    \includegraphics[width=\textwidth]{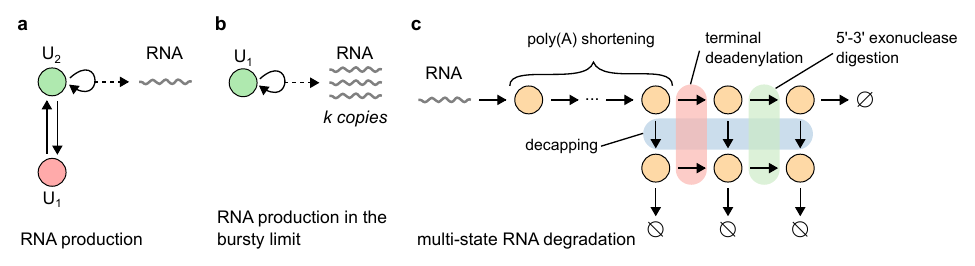}
    \caption{A stochastic gene expression model with multi-state RNA degradation. (a) RNA production described by the telegraph process, which is equivalent to the renewal process ($G$). (b) The bursty limit of the telegraph process, which is equivalent to the Poisson process with batch arrivals ($M^{X}$). (c) Multi-state model of eukaryotic RNA degradation consisting of poly(A) shortening, decapping, terminal deadenylation and 5'-3' exonuclease digestion. The model of RNA degradation has been adapted from~\cite{Cao_2001}.}
    \label{fig5}
\end{figure}

\subsection*{\texorpdfstring{$M^X/G/\infty$}{M\^X/G/infinity} queue}

The $M^X/G/\infty$ queue is an infinite-server queue in which the interarrival times are exponentially distributed, customers arrive in batches with a general (arbitrary) batch-size distribution, and the service times have a general (arbitrary) distribution. It is a special case of the $G^X/G/\infty$ queue when the arrivals are Poisson. Fig.~\ref{fig5} illustrates a stochastic gene expression model with a detailed, multi-state process of eukaryotic RNA degradation that can be analysed using the $M^X/G/\infty$ queue. In the limit that the gene spends most of its time in the off state, RNA production, as described by the telegraph process [Fig.~\ref{fig5}(a)], is replaced with the Poisson process with batch arrivals [Fig.~\ref{fig5}(b)], making the model equivalent to the $M^{X}/G/\infty$ queue. The RNA degradation process, adapted from Ref.~\cite{Cao_2001}, includes poly(A) shortening, decapping, terminal deadenylation and 5'-3' exonuclease digestion. Other examples of multi-state RNA processing are RNA splicing and nuclear export. In Ref.~\cite{Gorin_2022}, a model for RNA splicing has been proposed in which a parent RNA is produced in bursts, each of which then goes through a number of irreversible steps representing intron splicing and RNA degradation. A similar multi-state model has been considered for nascent RNA, in which multiple states represent positions of the RNA polymerase on the gene, i.e. the length of the nascent RNA~\cite{Filatova_2021,Szavits_2022}. This model has switching between two states, which in the bursty limit (when the off rate is large relative to the on rate) maps to the $M^X/G/\infty$ queue. A model that accounts for simple nuclear export was proposed in Ref.~\cite{Singh_2012}. This model describes bursty production of nuclear RNA ($M_n$), which is then transported to the cytoplasm ($M_c$) and is eventually degraded. For the total RNA $M$, $M=M_n+M_c$, this model maps to the $M^X/G/\infty$ queue in which RNA undergoes a two-step degradation process. 

The appeal of the $M^X/G/\infty$ queue is that it accounts for bursty expression and complex RNA degradation, while being mathematically tractable. If we assume that no RNA is present at the time $t=0$, then the probability generating function of the queue length distribution $P(m,t)$ at a later time $t$ is given by
\begin{equation}
    \label{Gt-MXGI}
    G(z,t)=\text{exp}\left\{-\rho\int_{0}^{t}dt' \left[1-A\left(z+(1-z)H(t')\right)\right]\right\},
\end{equation}
where $\rho$ is the arrival rate, $H(t)$ is the cumulative distribution function of the RNA degradation time, and $A(z)$ is the probability generating function of the batch-size distribution~\cite{Liu_1990}. From Eq.~(\ref{Gt-MXGI}), by knowing $H(t)$ and $A(z)$, one can compute the queue length distribution $P(m,t)$ and all its moments, see Appendix~\ref{appendix-c} for further details. The stationary limit is obtained by letting $t\rightarrow\infty$ in Eq.~(\ref{Gt-MXGI}).
Interestingly, when the batch sizes are fixed to $1$ (the $M/G/\infty$ queue), then the stationary queue length distribution is Poisson with rate parameter $\rho\beta$, where $\beta$ is the mean service time. Hence, when RNA production is limited to a single rate-limiting step, then the stationary RNA distribution cannot be used to distinguish between different mechanisms of RNA degradation. This result was first discussed in Ref.~\cite{Thattai_2016} as a possible explanation of why RNA number distributions of many genes in {\it E. coli} and {\it S. cerevisiae} are close to Poisson, even though RNA decay in those organisms is known to be a complex, multi-stage process.

Finally, it is worth noting that queues with batch arrivals can be also used to model stochasticity in protein numbers. This is since when mRNA ($M$) degrades much faster than protein ($P$), it can be shown ~\cite{shahrezaei2008analytical} that the standard model for the protein production process, $U_1\leftrightarrows U_2\xrightarrow[]{}U_2+M, M\xrightarrow[]{}\emptyset, M\xrightarrow[]{}M + P$, can be replaced by the effective reaction $G\xrightarrow[]{}G+k P$, where $k$ is a random variable distributed according to the geometric distribution. The perturbative approach of Ref.~\cite{shahrezaei2008analytical} can be extended to the case where there are more than two gene states, implying that if one is not interested in RNA fluctuations, then an effective bursty protein production process can always be derived as a reduced model valid under timescale separation conditions. In that sense, the $M^X/G/\infty$ queue can serve as a model for protein fluctuations where the degradation time distribution is arbitrary. If this distribution is exponential, then it is a crude model for protein dilution due to cell division~\cite{beentjes2020exact}; more complex distributions such as an Erlang distribution could describe the fact that multiple ubiquitination events are required before protein degradation~\cite{pedraza2008effects}. 

\subsection*{\texorpdfstring{$G/D/\infty$}{G/D/infinity} queue}

\begin{figure}[hbt]
    \centering
    \includegraphics[width=\textwidth]{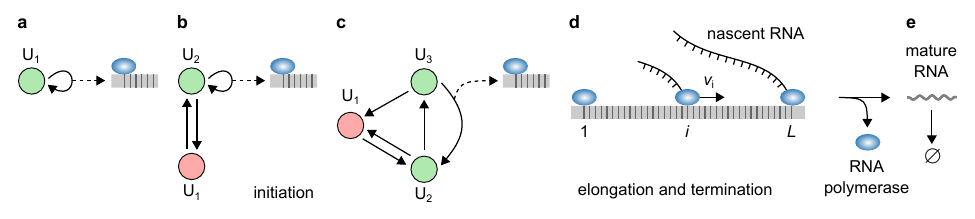}
    \caption{A stochastic gene expression model that consists of transcription initiation which produces nascent RNA, deterministic elongation and termination after which nascent RNA turns into mature RNA, and mature RNA degradation. (a)-(d) describe nascent RNA, which is equivalent to the $G/D/\infty$ queue. Transcription initiation can be any renewal process, for example, a MAP under the renewal condition (\ref{renewal-condition}). Examples of transcription initiation models include the one-state model of constitutive promoter (a)~\cite{Lafuerza_2011, Jiang_2021}, the two-state (telegraph) model of bursty promoter~\cite{Xu_2016} and (c) a three-state mechanistic model of bursty promoter that accounts for RNA polymerase recruitment and its release into productive elongation~\cite{Bartman_2019, Braichenko_2021}. (d) describes elongation and termination, which together take a fixed amount of time $T$ to finish. $L$ denotes the gene length, $v_i$ the RNA polymerase speed at position $i=1,\dots,L-1$, and $v_L$ the termination rate. The total time of elongation and termination is given by $T=\sum_{i=1}^{L}1/v_i$. (e) describes the turnover of mature RNA, which is equivalent to the $G/M/\infty$ queue. Since elongation and termination are deterministic, the arrival processes of nascent and mature RNA are the same.}
    \label{fig6}
\end{figure}

The $G/D/\infty$ queue is an infinite-server queue with independent and identically distributed interarrival times, and a deterministic service time. It is a special case of the $G^X/G/\infty$ queue. As discussed earlier, deterministic service time is a good approximation for stochastic service that consists of many steps, such that the service time distribution is sharply peaked at a non-zero value. An example of that scenario is transcription elongation, which consists of an RNA polymerase moving along the DNA one nucleotide at a time over thousands of nucleotides. Fig.~\ref{fig6} illustrates models of gene expression that account for multistep transcription initiation that produces nascent RNA [Figs.~(\ref{fig6}(a)-(c)], transcription elongation and termination after which nascent RNA turns into mature RNA [Fig.~\ref{fig6}(d)], and mature RNA degradation [Fig.~\ref{fig6}(e)]. In these examples, transcription initiation is modelled by a MAP under the renewal condition (\ref{renewal-condition}), whereas transcription elongation and termination are deterministic ($D$). Hence, the part of the model that describes nascent RNA is equivalent to the $G/D/\infty$ queue. Fig.~\ref{fig6}(a) shows the one-state model describing a gene that is always active~\cite{Lafuerza_2011, Jiang_2021}, Fig.~\ref{fig6}(b) shows the telegraph model describing a gene that switches between two states of activity and inactivity~\cite{Xu_2016, Fu_2022} and Fig.~\ref{fig6}(c) shows a three-state model that accounts for the binding of RNA polymerase at the promoter and its release into productive elongation~\cite{Bartman_2019,Braichenko_2021}. Other, more complicated models equivalent to the $PH/M/\infty$ queue, including the canonical model of transcription initiation in Fig.~\ref{fig4}(e), have been studied in Ref.~\cite{Szavits_2023}. 

We note that since elongation and termination in the models in Fig.~\ref{fig6} are deterministic, the interarrival times of nascent and mature RNA are equal. This in turn means that the mature RNA turnover is described by the $G/M/\infty$ queue with the same arrival process as the one describing the production of nascent RNA, whereas the service (RNA degradation) time is exponentially distributed. Consequently, the stationary mature RNA number distribution is independent of the elongation and termination time $T$, and can be determined using the results for the $G/M/\infty$ queue. We note that the full model of nascent and mature RNA is an example of two queues connected in series, such that the output of the first queue becomes an input of the second queue. Such queues are called tandem queues.

If we denote the constant service time by $T$, then the number of customers in a $G/D/\infty$ queue at any time $t$ is equal to the number of arrivals between $t-T$ ($0$ if $t<T$) and $t$. Hence, the queue length distribution of a $G/D/\infty$ queue can be obtained from the statistics of the number of arrivals in a given time interval, which in turn can be computed from the renewal theory~\cite{Cox_1967}. In the stationary limit, explicit results can be given for the Laplace transform of the queue length distribution with respect to the service time $T$,
\begin{equation}
    \label{LT-Pm-GDI}
    \mathcal{L}[P(m)](s)=\int_{0}^{\infty}dT P(m)e^{-sT}=\begin{dcases}
    \frac{\alpha s-1+\phi(s)}{\alpha s^2}, & m=0\\
    \frac{[1-\phi(s)]^2[\phi(s)]^{m-1}}{\alpha s^2},& m\geq 1.\end{dcases}
\end{equation}
where $\phi(s)$ is the Laplace transform of the interarrival time probability density function, and $\alpha$ is the mean interarrival time, $\alpha=-\phi'(0)$. On the other hand, moments of the stationary queue length distribution can be computed from the probability generating function $G(z)=\sum_{m=0}^{\infty}P(m)z^m$, whose Laplace transform with respect to $T$ is given by
\begin{equation}
    \label{LT-G-GDI}
    \mathcal{L}[G(z)](s)=\int_{0}^{\infty}dT G(z)e^{-sT}=\frac{1}{s}+\frac{(z-1)[1-\phi(s)]}{\alpha s^2[1-z\phi(s)]}.
\end{equation}
From here we get the following expression Fano factor $FF$
\begin{equation}
    FF=1+\frac{2}{T}\mathcal{L}^{-1}\left\{\frac{\phi(s)}{s^2(1-\phi(s))}\right\}-\frac{T}{\alpha},
\end{equation}
where $\mathcal{L}^{-1}$ denotes the inverse Laplace transform, which is here evaluated at $T$. Using this expression, it is possible to show that for any finite $T$, $FF\leq 1+2CV_{a}^{2}$, where $CV_{a}$ is the coefficient of variation of the interarrival time distribution, whereas for $T\rightarrow\infty$, the Fano factor approaches $CV_{a}^{2}$. The latter result has been previously derived for fluctuations in the number of cycles of a processive enzyme~\cite{Schnitzer_1995,Moffitt_2014}. If the arrival process is a MAP under the renewal condition (\ref{renewal-condition}), then the interarrival time distribution is a phase-type distribution whose Laplace transform $\phi(s)$ is a rational function of $s$. In this case, the queue length distribution $P(m)$ and its moments can be obtained from Eqs. (\ref{LT-Pm-GDI}) and (\ref{LT-G-GDI}), respectively, using partial fraction decomposition~\cite{Kung_1977}. In Appendix~\ref{appendix-d}, we demonstrate this procedure for the two-state model depicted in Fig.~\ref{fig6}(b), known as the delay telegraph model, which has been previously solved using the chemical master approach~\cite{Xu_2016,Fu_2022}.

The above results pertain to the stationary case. In the non-stationary case, the results are possible, but more involving. The non-stationary queue length distribution can be computed using renewal theory provided one can compute the distribution of the forward recurrence time (the time until the next arrival)~\cite{Cox_1967}, whereas the moments can be computed recursively without the knowledge of this distribution~\cite{Liu_1990}.  

\subsection*{\texorpdfstring{$MMPP/M/\infty$}{MMPP/M/infinity} queue}

The $MMPP/M/\infty$ queue is an infinite-server queue in which the arrival rate changes according to a finite-state Markov process, customers arrive one by one, and the service times are exponentially distributed. We note that an $MMPP$ is a special case of the $MAP$ in which the matrix $D_1$ describing arrivals is a diagonal matrix. This means that the state of the MAP does not change immediately upon arrival. If $D_1$ has all but one diagonal element equal to zero, then the arrival process is renewal and the results for the $G/M/\infty$ queue are applicable. Here we consider the general case in which the renewal condition (\ref{renewal-condition}) is not satisfied, i.e. in which $D_1$ is a diagonal matrix with more than one non-zero element on the diagonal. 

\begin{figure}[htb!]
    \centering
    \includegraphics[width=\textwidth]{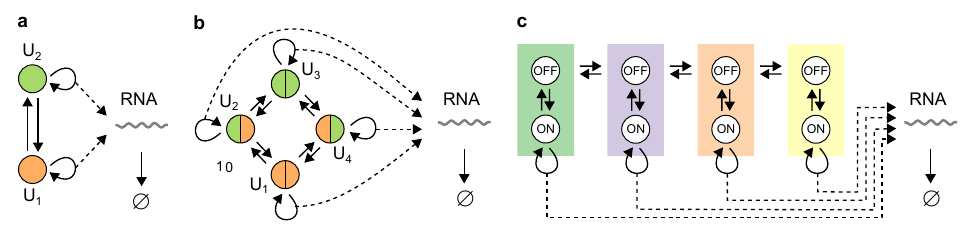}
    \caption{Examples of stochastic gene expression models that are equivalent to the $MMPP/M/\infty$ queue. (a) The leaky telegraph model in which the gene switches between two distinct microscopic transcription factor binding configurations, and produces RNA from both states~\cite{Kepler_2001}. (b) A generalization of the leaky telegraph model to include multiple transcription factor binding sites. Transitions are allowed only between configurations that differ by the presence (green colour) or absence (orange colour) of one transcription factor~\cite{Lammers_2020}. This particular model accounts for the binding of two transcription factors, yielding $2^2=4$ gene states. (c) A stochastic model describing mRNA production from lysogeny maintenance promoter of bacteriophage lambda. Different gene states correspond to different binding combinations of the lambda repressor Cl. Adapted from Ref.~\cite{Sepulveda_2016}.}
    \label{fig7}
\end{figure}

Unlike the models we discussed so far that all had a single active state, the $MMPP/M/\infty$ queue describes a gene that produces RNA from multiple active states. Switching between these states results in a variation of the transcription rate over time, in contrast to a single active state from which transcription occurs at a constant rate. Variability in the transcription rate is thought to occur due to multiple ways in which various molecules involved in transcription such as RNA polymerases, activators, repressors and inducers interact with gene regulatory elements such as promoters and enhancers. Examples of models with multiple active states and exponential RNA degradation are shown in Fig.~\ref{fig7}. Fig.~\ref{fig7}(a) shows the leaky telegraph model in which the gene switches between two distinct microscopic transcription factor binding configurations, and produces RNA from both configurations~\cite{Kepler_2001}. Typically, the leaky active state is responsible for the low (basal) transcription rate, whereas the other active state has a much higher transcription rate. Fig.~\ref{fig7}(b) shows a generalization of the leaky telegraph model to include multiple transcription factor binding sites, leading to multiple active states~\cite{Lammers_2020}. Fig.~\ref{fig7}(c) shows a model with four active states applied to mRNA production from lysogeny maintenance promoter of bacteriophage lambda, in which gene states correspond to different binding combinations of the lambda repressor Cl~\cite{Sepulveda_2016}. Other examples not shown here include: a model with two active states describing the induction of c-Fos transcription in response to p-ERK signalling in human cells~\cite{Munsky_2015}, a model with two active states describing expression of \textit{eve} stripe 2 in fruit fly (\textit{Drosophila})~\cite{Bothma_2014,Holloway_2017}, and a model with four active states describing \textit{STL1} expression in baker's yeast (\textit{Saccharomyces cerevisiae})~\cite{Neuert_2013,munsky2018distribution}.

Although having multiple active states compared to having only one may seem like a minor generalization, from the standpoint of queueing theory models with one active state and models with multiple active states are fundamentally different. Models with one active state have renewal arrivals, meaning that the production of RNA is determined solely by the interarrival time distribution. Consequently, the chemical master equation is needed only to compute the interarrival time distribution. On the other hand, models with multiple active states do not have a unique interrarrival time distribution. Instead, the interarrival time distribution changes depending on the active state an RNA molecule was produced from. For these type of models, we need the chemical master equation to track down the gene state changes. Since the chemical master equation is often too difficult to solve, the results for the $MMPP/M/\infty$ queue are limited to moments of the queue length distribution, compared to the $G/M/\infty$ queue for which the queue length distribution is also known. 

We present here the main results for the stationary case and leave the details, including the non-stationary case, for Appendix~\ref{appendix-e}. These results were first obtained in Ref.~\cite{Ocinneide_1986}. In the context of gene expression, similar results were obtained in Refs.~\cite{Sanchez_2008, Innocentini_2013, Ham_2020}. Let $X(t)$ denote the state of the arrival process (the gene state), $N(t)$ the queue length (the number of RNA) at time $t$, $P_{i}(m,t)$ the joint probability that $X(t)=i$ and $N(t)=m$, and $\bm{P}(m,t)$ the row matrix $(P_1(m,t),\dots,P_S(m,t))$. Let $\bm{f}_{s}(t) $ denote the $s$-th factorial moment of $\bm{P}(m,t)$,
\begin{equation}
    \label{fracmomt-MMPPMI}
    \bm{f}_{s}(t)=s!\sum_{m=s}^{\infty}\binom{m}{s}\bm{P}(m,t), \quad s=0,1,2,\dots.
\end{equation}
We note that the zeroth-order factorial moment $\bm{f}_0(t)=\sum_{m=0}^{\infty}\bm{P}(m,t)=\bm{P(t)}=(P_1(t),\dots,P_S(t))$, where $P_i(t)$ is the probability that the queue is in state $i$ at time $t$ regardless of the queue length. Let $P(m)=\lim_{t\rightarrow\infty}\bm{P}(m,t)$ and $\bm{f}_s=\lim_{t\rightarrow\infty}$ denote the queue length distribution and the $s$-th factorial moment in the stationary limit, respectively. Then $\bm{f}_{s}$ can be computed recursively from
\begin{equation}
    \label{fracmom-eq-MMPPMI}
    \bm{f}_s=s\bm{f}_{s-1}D_1(s\lambda I-D_0-D_1)^{-1},\quad s=1,2,3\dots,
\end{equation}
where $\lambda$ is the service (RNA degradation) rate, and $\bm{f}_0=\bm{P}$ satisfies the stationary master equation $\bm{P}(D_0+D_1)=0$. From here, all factorial moments $\bm{f}_s$ can be easily computed. Specifically, the Fano factor of the queue length distribution is given by
\begin{equation}
    FF=1+\frac{\bm{P}D_1(\lambda I-D)^{-1}(D_1\bm{1}^T)}{\bm{P}(D_1\bm{1}^T)}-\frac{\bm{P}(D_1\bm{1}^T)}{\lambda},
\end{equation}
where $\bm{1}^T$ is a $S\times 1$ matrix made of $1$s. The stationary probabilities $\bm{P}(m)$ can be computed recursively from
\begin{equation}
    \label{Pm-MMPPMI}
	\bm{P}(m-1)D_1-\bm{P}(m)(-D_0+m \lambda I)+(m+1)\lambda\bm{P}(m+1)=0,\quad m=0,1,2\dots,
\end{equation}
where $\bm{P}(-1)\equiv 0$, and $\bm{P}(0)$ can be computed from
\begin{equation}
    \label{P0-MMPPMI}
    \bm{P}(0)=\sum_{s=0}^{\infty}\frac{\bm{f}_s (-1)^s}{s!}.
\end{equation}
Based on these results, the procedure to compute the stationary queue length distribution is the following. In the first step, $\bm{f}_0=\bm{P}$ is computed by solving $\bm{P}(D_0+D_1)=0$. In the second step, the first $K$ factorial moments $\bm{f}_s$ are computed using Eq. (\ref{fracmom-eq-MMPPMI}), and $\bm{P}(0)$ is approximated by the sum of the first $K$ terms in Eq. (\ref{P0-MMPPMI}). The integer $K$ is selected to achieve the desired numerical precision of $\bm{P}(0)$. In the third step, $\bm{P}(m)$ is computed recursively using Eq. (\ref{Pm-MMPPMI}) up to some value of $m$ for which $P(m)$ becomes negligibly small. In the fourth and final step, $\bm{P}(m)$ is multiplied by $\bm{1}^T$ to get $P(m)=\sum_{i=1}^{S}P_{i}(m)$. This procedure can be implemented numerically~\cite{Ocinneide_1986}. 

\section*{Discussion}

As a theory that is more than a hundred years old, queueing theory is rich and vast. In this review, we focused on old, classical results in queueing theory that are directly applicable to RNA production and degradation models traditionally used in modelling gene expression, particularly in the setting of finite-state Markov processes describing discrete promoter states and multi-state RNA degradation pathways. Following this tradition, we set up a general model of RNA production as a Markovian arrival process, which assumes that all transitions are Markovian. However, we emphasize that queues with renewal arrivals, such as the $G/M/\infty$ and $G/D/\infty$ queues or more generally the $G^{X}/G/\infty$ queue, can be used to model gene expression beyond the Markovian framework, since in these queues the interarrival time distribution is arbitrary. An example of such model is a generalized telegraph model in which the time spent in the off state has an arbitrary probability distribution~\cite{Schwabe_2012}. This model has renewal arrivals, because the time spent in the on state is exponentially distributed, meaning that immediately after arrival the gene has no memory of how much time it has already spent in the on state. For this model, the Laplace transform of the interarrival distribution can be found and applied to the $G/M/\infty$ queue to get the stationary probability distribution of the RNA number, avoiding laborious derivation using the master equation approach~\cite{Shi_2020}. This example shows just how important are renewal arrivals in modelling gene expression: they account for uncorrelated interarrival times of any complexity, but retain the analytical tractability. Since the interarrival times between individual transcriptional events can now be measured experimentally~\cite{Muthukrishnan_2012,Tantale_2016,Douaihy_2023}, queueing theory can be used to include experimentally measured interarrival time distributions without resorting to their Markovian interpretation. 

Once we move away from renewal arrivals, there are many results potentially useful for gene expression modelling that we did not cover in detail. We first mention the $BMAP/G/\infty$ queue, where customers arrive in batches according to a batch Markovian arrival process (BMAP), and the service times are generally distributed. This queueing system describes our stochastic model of gene expression in Fig.~\ref{fig1} in the most general setting. Results for this queueing system are limited and quite complicated, however numerically feasible formulas have been derived for service times that are phase-type distributed~\cite{Masuyama_2002}. Another type of non-renewal processes which we mentioned only briefly are semi-Markov processes (SMP). Semi-Markov processes change their interarrival time distribution according to a finite-state Markov process. In that sense, they can be considered as Markov-modulated renewal processes. A generalization of the $G/M/\infty$ queue to semi-Markov arrivals is the $SMP/M/\infty$ queue, for which the stationary queue length distribution and the Laplace transform of the non-stationary queue length distribution have been computed in Ref.~\cite{Neuts_1972}. More general is the $SMP/G/\infty$ queue, in which the service time distribution is arbitrary. This queueing system was studied in Ref.~\cite{Liu_1991}, where recurrence relations for (binomial) moments of both non-stationary and stationary queue length distributions have been derived. We showed in Eq. (\ref{MAP-to-SMP}) that the MAP is a special case of the SMP. An advantage of the latter approach is that interarrival time distributions can be described by any suitable, user-defined function rather than a phase-type distribution as for a MAP. Practically, this means that the SMP description has fewer parameters than MAP. For example, a phase-type distribution of the hypoexponential type which is the distribution of a random variable composed of $N$ exponential distributions each with their own rate could be approximated by a two-parameter continuous distribution such as the gamma distribution. Hence, the SMP maybe useful as a reduced version of complex models of gene expression. 

One limitation of the general stochastic model for RNA production in Fig.~\ref{fig1} is that all its rate constants are assumed to be time-independent. This limitation can be addressed using non-stationary queueing systems. A classic queueing system in this regard is the $M_{t}/G/\infty$ queue, where the subscript $t$ denotes that the arrival rate (of the Poisson process) is time-dependent. It is well-known that the queue length distribution of the $M_{t}/G/\infty$ queue is Poisson distributed~\cite{Eick_1993}. Perhaps more interesting to gene expression modelling is the $M_{t}^{X}/G/\infty$ queue, where customers arrive in batches. The probability generating function of the queue length for this queueing system is also known exactly~\cite{Shanbhag_1966}. This result opens many possibilities for studying bursty gene expression (where expression occurs in intermittent bursts \cite{rodriguez2020transcription}) under time-dependent conditions. For example, it is known that the identities and intensities of different extracellular time-dependent signals are transmitted by modulation of certain transcription factors in the cytoplasm, which exert an influence on gene expression upon their translocation to the nucleus~\cite{hao2012signal,paszek2010oscillatory}. 

Finally, we mention two open problems in queueing theory that are relevant for gene expression modelling. The first problem is extending gene expression models to include both RNA and proteins, and finding their joint probability distribution. In this case, there are two queues, one that describes RNAs and the other that describes proteins. The difficulty is that the arrival rate of the second queue (the protein production rate) is dependent on the number of customers in the first queue (the RNA number). This problem is not standard in queueing theory. Systems with multiple queues are typically studied in a way that the output of one queue becomes the input of another. Here, however, customers arriving at the first queue leave the system after service, instead of being routed to the second queue. This problem has been addressed recently by several authors~\cite{Dean_2020,Dean_2022,Fralix_2023}. The second problem concerns finding joint queue length probability distributions for tandem queues, in which customers leaving one queue are routed to the next queue. In general, tandem queues are difficult to solve, with notable exceptions being tandems of $M/G/\infty$ and $M_t/G/\infty$ queues~\cite{Eick_1993}. An example of tandem queue in gene expression is nascent RNA turning into nuclear RNA, which is then transported to the cytoplasm where it becomes cytoplasmic RNA. An open problem here is to find the joint distribution of nascent, nuclear and cytoplasmic RNA which can be measured experimentally~\cite{munsky2018distribution,Fu_2022,weidemann2023minimal}.  

Concluding, we have shown how a wide variety of models of gene expression can be formulated in terms of queueing theory. We hope this review stimulates anyone interested in quantitative biology to use the tools of queueing theory to analytically study the stochastic properties of complex and biologically realistic models of gene expression.   

\section*{Acknowledgments}
This work was supported by a Leverhulme Trust research award (RPG-2020-327).

\appendix
\section{Results for the \texorpdfstring{$G^X/G/\infty$ queue}{G\^X/G/infinity queue}}
\label{appendix-a}
\setcounter{equation}{0}
\renewcommand\theequation{A\arabic{equation}}

The following summary of results is from Ref.~\cite{Liu_1991}. Let $f(t)$ and $h(t)$ denote the probability density functions of the interarrival and service times, respectively. The mean interarrival time $\alpha$ and the mean service time $\mu$ are assumed to be finite. Let $T_n$ denote the time of the $n$-th arrival (called arrival epoch), and $X_n$ the number of customers arriving at $T_n$. For simplicity, we assume that the initial time $t=0$ is an arrival epoch ($T_0=0$). The queue length at time $t$ is denoted by $N(t)$, and we assume that $N(0)=0$. The batch size distribution is denoted by $P(X=k)=a_k$, and the probability generating function of $a_k$ is denoted by $A(z)$,
\begin{equation}
    A(z)=\sum_{k=1}^{\infty}a_k z^k.
\end{equation}
From here we define the $k$-th factorial moment of $a_k$ as
\begin{equation}
    A_k=\left.\frac{d^kA(z)}{dz^k}\right\vert_{z=1}=\sum_{r=k}^{\infty}r(r-1)\dots(r-k+1)a_r.
\end{equation}
The mean batch size $A_1$ and the variance $A_2+A_1-A_{1}^2$ are assumed to be finite.

We are interested in the queue length distribution $P(m,t)$ (the probability that $N(t)=m$) and its moments. Let $G(z,t)$ denote the probability generating function of $N(t)$,
\begin{equation}
    G(z,t)=\sum_{m=0}^{\infty}P(m,t)z^m.
\end{equation}
The $r$-th binomial moment of $P(m,t)$ is defined as
\begin{equation}
    B_{r}(t)=\frac{1}{r!}\left.\frac{d^r G(z,t)}{dz^r}\right\vert_{z=1}=\sum_{m=r}^{\infty}\binom{m}{r}P(m,t),\quad r=0,1,2\dots, 
\end{equation}
To find $B_r(t)$, we first define the renewal function $R(t)$ as the expected value of the number of arrivals observed up to time $t$. From renewal theory~\cite{Cox_1967} it follows that
\begin{equation}
    R(t)=\sum_{n=1}^{\infty}K_n(t), \quad K_n(t)=\int_{0}^{t}dt'f^{*n}(t'),
\end{equation}
where $f^{*n}(t)$ is the $n$-fold convolution of $f(t)$. The Laplace transform of $R(t)$ is related to the Laplace transform of $f(t)$, denoted by $\phi(s)$,
\begin{equation}
    \mathcal{L}[R](s)=\frac{\phi(s)}{s[1-\phi(s)]}.
\end{equation}

Assuming $A_k\leq Q^{k}$ for all $k\geq 1$, where $Q$ is a constant, the $r$-th binomial moment $B_{r}(t)$ can be computed recursively from
\begin{equation}
    B_0(t)=1,\quad B_r(t)=\sum_{k=1}^{r}\frac{A_k}{k!}\int_{0}^{t}B_{r-k}(t-y)[1-H(t-y)]^{k}dR(y),\quad r=1,2,3,\dots,
\end{equation}
whereas the queue length distribution $P(m,t)$ is given by
\begin{equation}
    \label{Pm-GXGI}
    P(m,t)=\sum_{r=m}^{\infty}(-1)^{r-m}\binom{r}{m}B_r(t).
\end{equation}
In the stationary limit, $B_r=\lim_{t\rightarrow\infty}B_r(t)$ is given by
\begin{equation}
    B_r=\alpha\sum_{k=1}^{r}\frac{A_k}{k!}\int_{0}^{\infty}B_{r-k}(t)[1-H(t)]^{k}dt,\quad r=1,2,3,\dots,
\end{equation}
whereas $P(m)=\lim_{t\rightarrow\infty}P(m,t)$ is given by Eq.~(\ref{Pm-GXGI}) with $B_t(t)$ replaced by $B_r$.
If sufficient number of binomial moments can be computed recursively, then the queue length distribution can be approximated by truncating the series in Eq.~(\ref{Pm-GXGI}).

\section{Results for the \texorpdfstring{$G/M/\infty$ queue}{G/M/infinity queue}}
\label{appendix-b}
\setcounter{equation}{0}
\renewcommand\theequation{B\arabic{equation}}

Let $f(t)$ denote the probability density function of the interarrival time, and $F(t)$ its cumulative distribution function, $F(t)=\int_{0}^{t}f(t')dt'$. We denote by $\phi(s)$ the Laplace transform of the inter-arrival time distribution $f(t)$, 
\begin{equation}
    \phi(s)=\mathcal{L}[f](s)=\int_{0}^{\infty}dt e^{-st}f(t),
\end{equation}
and by $\alpha$ the mean inter-arrival time,
\begin{equation}
    \alpha=\int_{0}^{\infty}dt\;t f(t)=-\left.\frac{d}{ds}\phi(s)\right\vert_{s=0}.
\end{equation}
The cumulative distribution function of the service time is given by
\begin{equation}
    H(t)=1-e^{-\lambda t},
\end{equation}
where $\lambda$ is the service rate. Both $\alpha$ and $\lambda$ are assumed to be finite.

Let $N(t)$ denote the queue length (the number of RNA) at time $t$, $P(m,t)$ the probability that $N(t)=m$, and $G(z,t)$ the corresponding probability generating function,
\begin{equation}
    G(z,t)=\sum_{m=0}^{\infty}P(m,t)z^m.
\end{equation}
The initial time $t=0$ is assumed to be an arrival epoch and the initial queue length is $N(0)=0$. Under these assumptions, the probability generating function $G(z,t)$ satisfies an integral equation that is derived using the renewal property of the arrival process~\cite{Takacs_1958},
\begin{equation}
    \label{G-eq-GMI}
    G(z,t)=1-F(t)+\int_{0}^{t}dt'f(t')G(z,t-t')\big\{1+(z-1)[1-H(t-t')]\big\}.
\end{equation}
Eq. (\ref{G-eq-GMI}) can be solved by Laplace transform, yielding
\begin{equation}
    \label{LT-Gt-GMI}
    \mathcal{L}[G](z,s)=\int_{0}^{\infty}dte^{-st}G(z,t)=\frac{1}{s}+\sum_{j=1}^{\infty}\frac{(z-1)^j}{s+j\lambda}\prod_{i=0}^{j-1}\frac{\phi(s+i \lambda)}{1-\phi(s+i \lambda)},
\end{equation}
from where the following expression for the Laplace transform of $P(m,t)$ is obtained,
\begin{equation}
    \label{LT-Pt-GMI}
    \mathcal{L}[P](m,s)=\int_{0}^{\infty}dt e^{-st}P(m,t)=\begin{dcases}\frac{1}{s}-\sum_{j=1}^{\infty}\frac{(-1)^{j-1}}{s+j\lambda}\prod_{i=0}^{j-1}\frac{\phi(s+i \lambda)}{1-\phi(s+i \lambda)}, & m=0\\
    \sum_{j=m}^{\infty}\frac{(-1)^{j-m} \binom{j}{m}}{s+j\lambda}\prod_{i=0}^{j-1}\frac{\phi(s+i \lambda)}{1-\phi(s+i \lambda)}, & m\geq 1.
    \end{dcases}.
\end{equation}
The queue length distribution can be obtained from this expression by inverting the Laplace transform. If the arrival process is a renewal MAP, then $\phi(s)$ is a rational function of $s$ (a ratio of two polynomials), and the Laplace transform can be inverted using partial fraction decomposition~\cite{Kung_1977}. 

The stationary distribution $P(m)=\lim_{t\rightarrow\infty}P(m,t)$ can be obtained by taking the limit $\lim_{s\rightarrow 0}s\mathcal{L}[P(m,s)]$, which gives
\begin{equation}
    P(m)=\begin{dcases}
        1-\sum_{j=1}^{\infty}(-1)^{j-1}\frac{C_{j-1}}{\alpha \lambda j}, & m=0\\
        \sum_{j=m}^{\infty}(-1)^{j-m}\binom{j}{m}\frac{C_{j-1}}{\alpha \lambda j}, & m\geq 1
    \end{dcases},
\end{equation}
where $C_i$ for $i=0,1,\dots$ are defined as
\begin{equation}
    C_0=1,\quad C_j=\prod_{i=1}^{j}\frac{\phi(i\lambda)}{1-\phi(i \lambda)},\quad j\geq 1.
\end{equation}
The moments of $P(m)$ can be computed from the probability generating function $G(z)$, which is given by
\begin{equation}
    \label{G-GMI}
    G(z)=\sum_{m=0}^{\infty}P(m) z^m=1+\sum_{m=1}^{\infty}\frac{C_{m-1}}{\alpha\lambda m}(z-1)^{m}.
\end{equation}
The mean and the variance of the stationary queue length distribution read
\begin{equation}
    \mu=\frac{1}{\alpha\lambda}, \quad \sigma^{2}=\frac{1}{\alpha\lambda}\left[\frac{1}{1-\phi(\lambda)}-\frac{1}{\alpha\lambda}\right],
\end{equation}
where the expression in the square brackets is equal to the Fano factor $FF$,
\begin{equation}
    FF=\frac{\sigma^{2}}{\mu}=\frac{1}{1-\phi(\lambda)}-\frac{1}{\alpha\lambda}.
\end{equation}

The above result can be used to obtain the lower and upper bounds on the Fano factor for any distribution of the inter-arrival times with finite mean and variance. Let $CV_{a}$ denote the coefficient of variation of the inter-arrival time distribution and $\delta=CV_{a}^{2}$. The following sharp bounds on the Laplace transform $\phi(s)$ were derived in Ref.~\cite{Eckberg_1977},
\begin{equation}
    e^{-\alpha s}\leq\phi(s)\leq \frac{\delta}{1+\delta}+\frac{1}{1+\delta}e^{-\alpha s(1+\delta)},\quad s\geq 0.
\end{equation}
Applying this result to the Fano factor yields
\begin{equation}
    u(\alpha\lambda)\leq FF\leq \left(1+\delta\right)u\left(\alpha\lambda(1+\delta)\right),
\end{equation}
where $u(x)=1/(1-e^{-x})-1/x$. Since the function $u(x)$ is monotonically increasing from $u(0)=1/2$ to $u(\infty)=1$, we conclude that 
\begin{equation}
    \frac{1}{2}\leq FF\leq 1+CV_{a}^{2}.
\end{equation}
This result shows that the Fano factor of the queue length (the number of RNA) in the stationary limit cannot be smaller than $1/2$, regardless of the inter-arrival time distribution or the service (RNA degradation) rate.

Next, we use Eq. (\ref{G-GMI}) to derive the stationary probability generating function $G(z)$ for the stochastic models of gene expression with arbitrary connections between gene states (as in Fig.~\ref{fig1}) under the renewal condition (\ref{renewal-condition}) of the MAP. In this case, the inter-arrival time distribution is $PH(\bm{\kappa},D_0)$, whose Laplace transform $\phi(s)$ is given by
\begin{equation}
    \label{LT-f-GMI}
    \phi(s)=\bm{\kappa}\frac{1}{sI-D_0}(-D_0\bm{1}^T),
\end{equation}
where $I$ is the $S\times S$ identity matrix. As $\phi(s)$ is a rational function of $s$ and $\phi(0)=1$, we can always write $\phi(s)/[1-\phi(s)]$ in the following form,
\begin{equation}
    \label{LT-f-pars}
    \frac{\phi(s)}{1-\phi(s)}=\frac{c(s+a_1)\dots(s+a_p)}{s(s+b_1)\dots(s+b_q)},
\end{equation}
where we assumed that the two polynomials on the right-hand side are coprime. Eq.~(\ref{LT-f-pars}) serves as an implicit definition for the parameters $c$, $p$, $q$, $c$, $a_1,\dots,a_p$ and $b_1,\dots,b_q$. We note that since $\phi(s)=1-\alpha s+O(s^2)$ as $s\rightarrow 0$,  
\begin{equation}
    \alpha=\frac{1}{c}\left(\frac{b_1\dots b_q}{a_1\dots a_p}\right).
\end{equation}
Inserting (\ref{LT-f-pars}) into (\ref{G-GMI}) gives, after some algebra, a remarkably compact result
\begin{equation}
    \label{G-PHMI-app}
    G(z)={}_p F_{q}\left(\frac{a_1}{\lambda},\dots,\frac{a_p}{\lambda};\frac{b_1}{\lambda},\dots,\frac{b_q}{\lambda};c \lambda^{p-q-1}\left(z-1\right)\right),
\end{equation}
where ${}_p F_{q}(a_1,\dots,a_p;b_1,\dots,b_q;z)$ is the generalized hypergeometric function defined as
\begin{equation}
    {}_p F_{q}(a_1,\dots,a_p;b_1,\dots,b_q;z)=\sum_{m=0}^{\infty}\frac{(a_1)_{m}\dots(a_p)_{m}}{(b_1)_{m}\dots(b_q)_{m}}\frac{z^m}{m!},
\end{equation}
and $(x)_{n}=x(x+1)\dots(x+n-1)$ is the rising factorial. The queue length distribution $P(m)$ is obtained by taking the $m$-th derivative of $G(z)$ evaluated at $z=0$, which yields
\begin{equation}
    \label{Pm-PHMI}
    P(m)=\frac{(c\lambda^{p-q-1})^m}{m!}\frac{(a_1/\lambda)_m\dots(a_p/\lambda)_m}{(b_1/\lambda)_m\dots(b_q/\lambda)_m}\;{}_p F_{q}\left(\frac{a_1}{\lambda}+m,\dots,\frac{a_p}{\lambda}+m;\frac{b_1}{\lambda}+m,\dots,\frac{b_q}{\lambda}+m;-c\lambda^{p-q-1}\right).
\end{equation}

As an example of how easy it is to solve gene expression models this way, we show here how to compute the stationary RNA number distribution for the telegraph model in which the gene switches between two states $U_1$ and $U_2$ with rates $W_0(1\rightarrow 2)=k_{\text{on}}$ and $W_0(2\rightarrow 1)=k_{\text{off}}$, and produces RNA from the state $U_2$ with rate $W_1(2\rightarrow 2)=k_{\text{syn}}$ [Fig.~\ref{fig4}(b)]. The reaction scheme for this model is
\begin{equation}
    U_1\xrightleftharpoons[k_{\text{off}}]{k_{\text{on}}}U_2\xrightarrow[]{k_{\text{syn}}}U_2+M,\quad M\xrightarrow[]{\lambda}\emptyset.
    \label{telegraph-model}
\end{equation}
The Laplace transform of the inter-arrival time distribution for the telegraph model is given by~\cite{Szavits_2023}
\begin{equation}
    \label{LT-f-telegraph}
    \phi(s)=\frac{k_{\text{syn}}(k_{\text{on}}+s)}{s^2+s(k_{\text{on}}+k_{\text{off}}+k_{\text{syn}})+k_{\text{on}}k_{\text{syn}}}.
\end{equation}
After inserting Eq. (\ref{LT-f-telegraph}) into $\phi(s)/[1-\phi(s)]$, we get
\begin{equation}
    \frac{\phi(s)}{1-\phi(s)}=\frac{k_{\text{syn}}(s+k_{\text{on}})}{s(s+k_{\text{on}}+k_{\text{off}})}.
\end{equation}
By comparing this expression to Eq. (\ref{LT-f-pars}), we identify $p=1$, $q=1$, $c=k_{\text{syn}}$, $a_1=k_{\text{on}}$ and $b_1=k_{\text{on}}+k_{\text{off}}$. After inserting these results into Eq. (\ref{G-PHMI-app}), we get the following expression for the probability generating function $G(z)$,
\begin{equation}
    \label{G-telegraph}
    G(z)={}_{1}F_{1}\left(\frac{k_{\text{on}}}{\lambda};\frac{k_{\text{on}}}{\lambda}+\frac{k_{\text{off}}}{\lambda};\frac{k_{\text{syn}}}{\lambda}(z-1)\right).
\end{equation}
We note that the process of switching between two states with arrivals from one of the states is known in queueing theory as the interrupted Poisson process (IPP). The telegraph model is therefore equivalent to the $IPP/M/\infty$ queue. The steady-state distribution of the queue length for this queueing system was first derived in 1973~\cite{Kuczura_1973}, more than 20 years before the seminal paper by Peccoud and Ycart on the telegraph model~\cite{Peccoud_1995}.

\section{Results for the \texorpdfstring{$M^X/G/\infty$ queue}{M\^X/G/infinity queue}}
\label{appendix-c}
\setcounter{equation}{0}
\renewcommand\theequation{C\arabic{equation}}

Let $\rho$ denote the arrival rate, $X\geq 1$ the batch size, and $P(X=k)=a_k$ the batch size distribution. The cumulative distribution of the service time is denoted by $H(t)$. We denote by $A(z)$ the probability generating function of $X$,
\begin{equation}
    \label{A-MXGI}
    A(z)=\sum_{k=0}^{\infty}a_k z^k.
\end{equation}
For any positive integer $n$, we define the $n$-th factorial moment of $X$ as
\begin{equation}
    A_n=\left.\frac{d^n}{dz^n}A(z)\right\vert_{z=1}=\sum_{k=n}^{\infty}k(k-1)\dots(k-n+1)a_k.
\end{equation}
The mean and the variance of $X$ are assumed to be finite. 

Let $N(t)$ denote the queue length at time $t$. The initial time is chosen to be an arrival epoch, and the system is initially empty, $N(0)=0$. Under these assumptions, the probability generating function $G(z,t)$ of the queue length $N(t)$ reads~\cite{Liu_1990}
\begin{equation}
    \label{Gt-MXGI-appendix}
    G(z,t)=\text{exp}\left\{-\rho\int_{0}^{t}dt' \left[1-A\left(z+(1-z)H(t')\right)\right]\right\}.
\end{equation}
From here, we get the following expressions for the mean and the variance of the queue length at time $t$,
\begin{subequations}
\begin{align}
    \label{mean-MXGI}
    & \mu(t)=\rho A_1\int_{0}^{t}dt' \left[1-H(t')\right],\\
    \label{var-MXGI}
    & \sigma^{2}(t)=\mu(t)+\rho A_2\int_{0}^{t}dt'\left[1-H(t')\right]^2.
\end{align}    
\end{subequations}
The stationary limit is obtained by letting $t\rightarrow\infty$ in Eq. (\ref{Gt-MXGI-appendix}), (\ref{mean-MXGI}) and (\ref{var-MXGI}), i.e. by replacing $\int_{0}^{t}dt'$ with $\int_{0}^{\infty}dt'$.

As an example, we consider arrivals whose batch size $X$ is geometrically distributed,
\begin{equation}
    \label{Px-MXGI}
    P(X=k)=(1-p)p^{k}, \quad k=0,1,\dots.   
\end{equation}
This distribution can be derived from the reaction scheme of the telegraph model in Eq. (\ref{telegraph-model}) when the gene spends most of its time in the inactive state ($k_{\text{off}}\gg k_{\text{on}}$). In that case, RNA synthesis can be described by the effective reaction 
\begin{equation}
    G\xrightarrow[]{k_{\text{on}}}G+kM,   
\end{equation}
where $k$ follows the geometric distribution in Eq. (\ref{Px-MXGI}) with $p=k_{\text{syn}}/(k_{\text{off}}+k_{\text{syn}})$. The mean burst size for this distribution is $k_{\text{syn}}/k_{\text{off}}$. Inserting Eq. (\ref{Px-MXGI}) into Eq. (\ref{A-MXGI}), we get
\begin{equation}
    A(z)=\frac{1-p}{1-pz}.
\end{equation}

We first consider one-step degradation, which is equivalent to the $M^{X}/M/\infty$ queue. In this case, $H(t)=1-e^{-\lambda t}$, where $\lambda$ is the RNA degradation rate. Inserting $A(z)$ and $H(t)$ into Eq. (\ref{Gt-MXGI}) and taking the stationary limit $t\rightarrow\infty$ yields
\begin{equation}
    \label{G-example1-MXGI}
    G(z)=\left[\frac{1-p}{1-pz}\right]^{r},
\end{equation}
where $r=k_{\text{on}}/\lambda$. From here, expanding $G(z)$ around $z=0$ gives  
\begin{equation}
    P(m)=\binom{m+r-1}{m}p^{m}(1-p)^{r},\quad p=\frac{k_{\text{syn}}}{k_{\text{off}}+k_{\text{syn}}},\quad r=\frac{k_{\text{on}}}{\lambda}.
\end{equation}
which is the negative binomial distribution $\text{NB}(r,1-p)$ that is often used in the analysis of single-cell data~\cite{Gruen_2014,Vallejos_2015}.

Next, we consider the reaction scheme
\begin{equation}
    \label{bursty-model-export}
    G\xrightarrow[]{k_{\text{on}}}G+kM_{n},\quad M_{n}\xrightarrow[]{\lambda_e}M_c\xrightarrow[]{\lambda_c}\emptyset.   
\end{equation}
This model describes production of nuclear RNA $M_n$, which is transported to the cytoplasm where it becomes cytoplasmic RNA $M_c$. The model was studied in Ref.~\cite{Singh_2012}, where the stationary joint probability distribution $P(m_n,m_c)$ of the nuclear RNA number $m_n$ and cytoplasmic RNA number $m_c$ was considered. Using the master equation approach, the following expression for the probability generating function of nuclear and cytoplasmic RNA numbers was obtained,
\begin{equation}
    \label{Gxy-MXGI}
    G(x,y)=\sum_{m_n=0}^{\infty}\sum_{m_c=0}^{\infty}P(m_n,m_c)x^{m_n}y^{m_c}=\text{exp}\left\{-k_{\text{on}}\int_{0}^{\infty}dt[1-A(u(x,y,t))],\right\}
\end{equation}
where $A$ is the probability generating function of the batch size, and $u$ is given by
\begin{equation}
    u(x,y,t)=1+\left[(x-1)+\frac{(y-1)\lambda_e}{\lambda_c-\lambda_e}\right]e^{-\lambda_e t}-\frac{(y-1)\lambda_e}{\lambda_c-\lambda_e}e^{-\lambda_c t}.
\end{equation}
The probability generating function of nuclear RNA is obtained by setting $y=1$, which is equivalent to the $M^{X}/M/\infty$ queue for which the probability generating function $G(z)$ is given by Eq. (\ref{G-example1-MXGI}) with $\lambda=\lambda_e$. On the other hand, the probability generating function of total (mature) RNA consisting of both nuclear and cytoplasmic RNA is obtained by setting $x=y$. This is equivalent to the $M^X/G/\infty$ queue, where the service time is the total time of nuclear export and cytoplasmic RNA degradation. This time is distributed according to the hypoexponential distribution
\begin{equation}
    \label{H2-MXGI}
    H(t)=1-\frac{\lambda_c}{\lambda_c-\lambda_e}e^{-\lambda_e t}+\frac{\lambda_e}{\lambda_c-\lambda_e}e^{-\lambda_c t}.
\end{equation}
Indeed, $G(x,x)$ obtained by setting $x=y$ in $u(x,y,t)$ is the same as $G(z)$ obtained by inserting $H(t)$ given by Eq. (\ref{H2-MXGI}) into Eq. (\ref{Gt-MXGI}) in the stationary limit $t\rightarrow\infty$. 

\setcounter{equation}{0}
\renewcommand\theequation{D\arabic{equation}}

\section{Results for the \texorpdfstring{$G/D/\infty$ queue}{G/D/infinity queue}}
\label{appendix-d}
\setcounter{equation}{0}
\renewcommand\theequation{D\arabic{equation}}

Let $T_n$ denote the time of the $n$-th arrival and $t_n$ the inter-arrival time between the $(n-1)$-th and $n$-th arrival, $t_n=T_{n}-T_{n-1}$. The initial time $t=0$ is assumed to be an arrival epoch, i.e. $T_0=0$. All interarrival times are taken from the same distribution whose probability density function is denoted by $f(t)$. Let $Y(t)$ denote the number of arrivals until time $t$. The probability that $Y(t)\geq n$ is given by
\begin{equation}
    P(Y(t)\geq n)=P(T_n\leq t)\equiv K_n(t)=\int_{0}^{t}dt'f^{*n}(t'),
\end{equation}
where $f^{*n}(t)$ is the $n$-fold convolution of $f(t)$. Since $P(Y(t)\geq n)=P(Y(t)=n)+P(Y(t)\geq n+1)$, we get 
\begin{equation}
    P(Y(t)=n)=K_{n}(t)-K_{n+1}(t).
\end{equation}
The mean number of arrivals in $(0,t)$, which is called the renewal function, is given by
\begin{equation}
    \label{RF-GDI}
    R(t)=\sum_{n=0}^{\infty}n P(Y(t)=n)=\sum_{n=1}^{\infty}K_n(t).
\end{equation}
The renewal function $R(t)$ can also be computed by inverting its Laplace transform, 
\begin{equation}
    \label{LT-RF-GDI}
    \mathcal{L}[R](s)=\frac{\phi(s)}{s[1-\phi(s)]},
\end{equation}
where $\phi(s)=\mathcal{L}[f](s)$. Let $T$ denote the service time. Since the service time is fixed, the queue length at time $t$, asuming $N(0)=0$, is equal to
\begin{equation}
    N(t)=\begin{cases}
    Y(t), & t\leq T,\\
    Y(t)-Y(t-T), & t>T.
    \end{cases}
\end{equation}
For $t\leq T$, the probability distribution $P(N(t)=m)\equiv P(m,t)=K_{m}(t)-K_{m+1}(t)$. For $t>T$, we introduce the forward recurrence time $\tau$, which is the time until the next arrival measured from some reference time $t_0$. The probability density function of $\tau$ can be computed from 
\begin{equation}
    \label{FRT-GDI}
    f_{t_0}(\tau)=f(t_0+\tau)+\int_{0}^{t_0}dt'r(t_0-t')f(t'+\tau),
\end{equation}
where $r(t)=dR/dt$ is called the renewal density, since $r(t)dt$ is equal to the probability that an arrival occurs in $(t,t+dt)$. The first term comes from having no arrivals before $t_0$, whereas the second term comes from having the previous arrival at some earlier time $t_0-t'$. If we know $f_{t_0}(\tau)$, then the non-stationary probability $P(m,t)$ can be computed as follows. For $m=0$, $P(0,t)$ is equal to the probability that the forward recurrence time $\tau$ is greater than $T$. For $m\geq 1$, $P(m,t)$ is equal to the convolution of $f_{t_0}$ and $K_{m-1}-K_m$ for $t_0=t-T$. Altogether, 
\begin{equation}
    \label{Pmt-GDI}
    P(m,t)=\begin{dcases}
    K_{m}(t)-K_{m+1}(t), & t\leq T, m\geq 0,\\
    \int_{T}^{\infty}d\tau f_{t-T}(\tau), & t>T,\;m=0,\\
    \int_{0}^{T}d\tau f_{t-T}(\tau)[K_{m-1}(T-\tau)-K_m(T-\tau)], & t>T,\; m\geq 1.
    \end{dcases}
\end{equation}
The moments of the non-stationary queue length distribution can be computed recursively without computing the forward recurrence time distribution, see Ref.~\cite{Liu_1990} for more details. The mean and the variance of the queue length read
\begin{subequations}
    \begin{align}
    \mu(t)&=\begin{dcases}
    R(t), & t\leq T,\\
    R(t)-R(t-T), & t>T,
    \end{dcases}\\
    \sigma^{2}(t)&=\begin{dcases}
    2\int_{0}^{t}dt'r(t')R(t-t')+R(t)-[R(t)]^2, & t\leq T,\\
    2\int_{t-T}^{t}dt'r(t')R(t-t')+\mu(t)-[\mu(t)]^2, & t>T.
    \end{dcases}
\end{align}
\end{subequations}

The above calculation simplifies in the stationary limit. In this limit, $\text{lim}_{t\rightarrow\infty}q(t)=1/\alpha$, where $\alpha$ is the mean inter-arrival time. Assuming that $\text{lim}_{t_0\rightarrow\infty}f_{t_0}(t_0+\tau)=0$, from Eq. (\ref{FRT-GDI}) it follows that
\begin{equation}
    \label{FRss-GDI}
    \lim_{t_0\rightarrow\infty}f_{t_0}(\tau)\equiv f_{\infty}(\tau)=\frac{1-F(\tau)}{\alpha},
\end{equation}
where $F(t)=\int_{0}^{t}dt'f(t')$ is the cumulative distribution function of the inter-arrival time. Inserting Eq. (\ref{FRss-GDI}) into Eq. (\ref{Pmt-GDI}) yields in the limit $t_{0}\rightarrow\infty$
\begin{equation}
    \label{Pm-GDI}
    P(m)=\begin{dcases}
    \int_{T}^{\infty}d\tau f_{\infty}(\tau), & m=0,\\
    \int_{0}^{T}d\tau f_{\infty}(\tau)[K_{m-1}(T-\tau)-K_m(T-\tau)], & m\geq 1.
    \end{dcases}
\end{equation}
By taking the Laplace transform of $P(m)$ with respect to $T$, we get
\begin{equation}
    \label{LT-Pm-GDI-app}
    \mathcal{L}[P(m)](s)=\int_{0}^{\infty}dT P(m)e^{-sT}=\begin{dcases}
    \frac{\alpha s-1+\phi(s)}{\alpha s^2}, & m=0\\
    \frac{[1-\phi(s)]^2[\phi(s)]^{m-1}}{\alpha s^2},& m\geq 1.\end{dcases}
\end{equation}
If the arrival process is a MAP under the renewal condition (\ref{renewal-condition}), then the inter-arrival time distribution is a phase-type distribution whose Laplace transform is a rational function of $s$. In this case, $P(m)$ can be obtained from Eq. (\ref{LT-Pm-GDI-app}) using partial fraction decomposition~\cite{Kung_1977}. The moments of $P(m)$ can be computed from the probability generating function $G(z)$ defined as
\begin{equation}
    G(z)=\sum_{m=0}^{\infty}z^m P(m).
\end{equation}
The Laplace transform of $G(z)$ with respect to $T$ is given by
\begin{equation}
    \label{LT-G-GDI-app}
    \mathcal{L}[G(z)](s)=\int_{0}^{\infty}dT G(z)e^{-sT}=\frac{1}{s}+\frac{(z-1)[1-\phi(s)]}{\alpha s^2[1-z\phi(s)]}.
\end{equation}
The mean and the variance of the queue length are given by
\begin{subequations}
\begin{align}
    \label{mean-GDI}
    & \mu=\frac{T}{\alpha},\\
    \label{var-GDI}
    & \sigma^2=\mathcal{L}^{-1}\left\{\frac{1+\phi(s)}{\alpha s^2[1-\phi(s)]}\right\}(T)-\left(\frac{T}{\alpha}\right)^2,
\end{align}
\end{subequations}
where $\mathcal{L}^{-1}\{\dots\}(T)$ is the inverse Laplace transform evaluated at $T$. 

From the above result, we obtain two general results on the Fano factor $FF$ without specifying the inter-arrival time distribution. The first result concerns the limit $T\rightarrow\infty$. By expanding $\phi(s)$ in Eq. (\ref{var-GDI}) around $s=0$ and collecting the lowest-order term, we get
\begin{equation}
    \label{limit1-GDI}
    \lim_{T\rightarrow\infty}FF=CV_{a}^{2},
\end{equation}
where $CV_{a}^{2}$ is the coefficient of variation of the inter-arrival time distribution. This result has been previously derived for fluctuations in the number of cycles of a processive enzyme~\cite{Schnitzer_1995,Moffitt_2014}. The second result gives an upper bound on the Fano factor in terms $CV_{a}^{2}$. By rearranging $1+\phi(s)$ into $1-\phi(s)+2\phi(s)$ and using Eq. (\ref{LT-RF-GDI}), we get
\begin{equation}
    \label{var2-GDI}
    \sigma^2=\frac{2}{\alpha}\int_{0}^{T}dt R(t)+\frac{T}{\alpha}-\left(\frac{T}{\alpha}\right)^2.
\end{equation}
The Fano factor $FF$ of the queue length is given by
\begin{equation}
    FF=\frac{\sigma^2}{\mu}=1+\frac{2}{T}\int_{0}^{T}dt R(t)-\frac{T}{\alpha}.
\end{equation}
This expression was derived in Ref.~\cite{Liu_1990} where the $G/D/\infty$ queue was considered as a special case of the $G^{X}/G/\infty$ queue. Using a general upper bound on the renewal function derived in Ref.~\cite{Lorden_1970}, $R(t)\leq t/\alpha+CV_{a}^{2}$, we get
\begin{equation}
    \label{limit2-GDI}
    FF\leq 1+2CV_{a}^{2}.
\end{equation}

As an example, we show how the stationary probability generating function $G_{nc}(z)$ of the nascent RNA number for the two-state model in Fig.~\ref{fig6}(b) can be easily computed from Eq. (\ref{LT-G-GDI-app}). The reaction scheme for this model is
\begin{equation}
    \label{delay-telegraph-model}
    U_1\xrightleftharpoons[k_{\text{off}}]{k_{\text{on}}}U_2\xrightarrow[]{k_{\text{syn}}}U_2+M_{nc},\quad M_{nc}\xRightarrow[]{T}M,\quad M\xrightarrow[]{\lambda}\emptyset,
\end{equation}
where $M_{nc}$ denotes nascent RNA and $\Rightarrow$ denotes a deterministic reaction that takes a fixed amount of time to finish. The Laplace transform of the inter-arrival time distribution for the two-state process is given by Eq. (\ref{LT-f-telegraph}). Inserting this result into Eq. (\ref{LT-G-GDI}), we get
\begin{equation}
    \mathcal{L}[G_{nc}(z)](s)=\frac{s+k_{\text{off}}+k_{\text{on}}-k_{\text{syn}} u+u/\alpha}{s^2+s(k_{\text{off}}+k_{\text{on}}-k_{\text{syn}} u)-k_{\text{on}} k_{\text{syn}} u},
\end{equation}
where $u=z-1$ and $\alpha=(k_{\text{on}}+k_{\text{off}})/(k_{\text{on}}k_{\text{syn}})$. This expression can be inverted using partial fraction decomposition as follows. In the first step, we find $\lambda_{1,2}$ such that $(s+\lambda_1)(s+\lambda_2)=s^2+s(k_{\text{off}}+k_{\text{on}}-k_{\text{syn}} u)-k_{\text{on}} k_{\text{syn}} u$, which gives
\begin{equation}
    \lambda_{1,2}=\frac{k_{\text{off}}+k_{\text{on}}-k_{\text{syn}} u\pm\sqrt{\Delta(u)}}{2},
\end{equation}
where $\Delta(u)=(k_{\text{off}}+k_{\text{on}}-k_{\text{syn}} u)^2+4k_{\text{on}} k_{\text{syn}}$. In the second step, we compute $A$ and $B$ such that
\begin{equation}
    \label{LT-G2-GDI}
    \mathcal{L}[G(z)](s)=\frac{A}{s+\lambda_1}+\frac{B}{s+\lambda_2},
\end{equation}
which gives
\begin{subequations}
    \begin{align}
        & A=\frac{-(k_{\text{on}}+k_{\text{off}})^2+\sqrt{\Delta}(k_{\text{on}}+k_{\text{off}})-(k_{\text{on}}-k_{\text{off}})k_{\text{syn}}u}{2\sqrt{\Delta}(k_{\text{on}}+k_{\text{off}})},\\
        & B=\frac{(k_{\text{on}}+k_{\text{off}})^2+\sqrt{\Delta}(k_{\text{on}}+k_{\text{off}})+(k_{\text{on}}-k_{\text{off}})k_{\text{syn}}u}{2\sqrt{\Delta}(k_{\text{on}}+k_{\text{off}})}.
\end{align}
\end{subequations}
Inserting $\lambda_{1}$, $\lambda_2$, $A$ and $B$ into (\ref{LT-G2-GDI}), and using $\mathcal{L}^{-1}[1/(s+\lambda)](T)=e^{-\lambda T}$ for $T\geq 0$, we get
\begin{align}
    G_{nc}(u)=&\frac{e^{-\lambda_1(u)T}}{2(k_{\text{on}}+k_{\text{off}})\sqrt{\Delta(u)}}\left\{(k_{\text{on}}+k_{\text{off}})\left[\sqrt{\Delta(u)}-(k_{\text{on}}+k_{\text{off}})\right]-(k_{\text{on}}-k_{\text{off}})k_{\text{syn}} u\right.\nonumber\\
    &+\left.(k_{\text{on}}+k_{\text{off}})\left[\sqrt{\Delta(u)}+(k_{\text{on}}+k_{\text{off}})\right]e^{\sqrt{\Delta(u)}T}+(k_{\text{on}}-k_{\text{off}})k_{\text{syn}} u\; e^{\sqrt{\Delta(u)}T}\right\},
    \label{G-telegraph-GDI}
\end{align}
The mean and the variance of the nascent RNA number are equal to
\begin{subequations}
   \begin{align}
    & \mu_{nc}=\frac{k_{\text{syn}}k_{\text{on}}T}{k_{\text{on}}+k_{\text{off}}},\\
    &\sigma_{nc}^{2}=\mu_{nc}\left\{1+\frac{2k_{\text{syn}} k_{\text{off}}}{T(k_{\text{on}}+k_{\text{off}})^3}\left[e^{-(k_{\text{on}}+k_{\text{off}})T}-1+(k_{\text{on}}+k_{\text{off}})T\right]\right\},
\end{align} 
\end{subequations}
where the expression in the curly brackets is the Fano factor of the nascent RNA number. The results for $G_{nc}(z)$, $\mu_{nc}$ and $\sigma_{nc}^{2}$ have been previously derived using the master equation approach~\cite{Xu_2016,Fu_2022}. Since we have switching between two states, it can easily be shown that
\begin{equation}
    CV_{a}^{2}=1+\frac{2k_{\text{sync}}k_{\text{off}}}{(k_{\text{on}}+k_{\text{off}})^2}.
\end{equation}
We see that the limit $\lim_{T\rightarrow\infty}FF_{nc}=CV_{a}^{2}$ in Eq. (\ref{limit1-GDI}) is satisfied, and so is the inequality in Eq. (\ref{limit2-GDI}), since $e^{-x}\leq 1$ for any $x\geq 0$.

\section{Results for the \texorpdfstring{$MMPP/M/\infty$ queue}{MMPP/M/infinity queue}}
\label{appendix-e}
\setcounter{equation}{0}
\renewcommand\theequation{E\arabic{equation}}

Let $X(t)$ denote the state of the arrival process (the gene state), $N(t)$ the queue length (the number of RNA) at time $t$, and $P_{i}(m,t)$ the joint probability that $X(t)=i$ and $N(t)=m$. Using matrices $D_0$ and $D_1$, the master equation for $\bm{P}(m,t)=(P_1(m,t),\dots,P_S(m,t))$ can be written as 
\begin{equation}
    \label{Pmt-MMPPMI}
	\frac{d}{dt}\bm{P}(m,t)=\bm{P}(m-1)D_1-\bm{P}(m)(-D_0+m \lambda I)+(m+1)\lambda\bm{P}(m+1),\quad m=0,1,2\dots,
\end{equation}
where $\bm{P}(-1)\equiv (0,\dots,0)$ $I$ is the $S\times S$ identity matrix and $\lambda$ is the RNA degradation rate. Let $\bm{f}_s(t)$ denote the $s$-th (vector) factorial moment of $\bm{P}(m,t)$,
\begin{equation}
    \label{fracmomt-MMPPMI-app}
    \bm{f}_{s}(t)=s!\sum_{m=s}^{\infty}\binom{m}{s}\bm{P}(m,t), \quad s=0,1,2,\dots.
\end{equation}
If $\bm{f}_s(0)$ exists for $s\geq 0$, then $\bm{f}_{s}(t)$ exists for all $s\geq 0$ and $t\geq 0$~\cite{Ocinneide_1986}. Furthermore, $\bm{f}_s(t)$ satisfy 
\begin{equation}
    \frac{d}{dt}\bm{f}_{s}(t)=s\bm{f}_{s-1}(t)D_1-\bm{f}_{s}(t)(s\lambda I-D_0-D_1),
\end{equation}
where $\bm{f}_{-1}\equiv 0$, which can be solved numerically to obtain non-stationary moments of the queue length distribution.

Let $\bm{P}(m)$ denote the stationary limit of $\bm{P}(m,t)$. In this limit, Eq. (\ref{Pmt-MMPPMI}) becomes
\begin{equation}
    \label{Pm-MMPPMI-app}
	\bm{P}(m-1)D_1-\bm{P}(m)(-D_0+m \lambda I)+(m+1)\lambda\bm{P}(m+1)=0,\quad m=0,1,2\dots,
\end{equation}
This equation can be seen as a recurrence relation for $\bm{P}(m)$, provided $\bm{P}(0)$ is known. The latter can be computed from 
\begin{equation}
    \label{P0-MMPPMI-app}
    \bm{P}(0)=\sum_{s=0}^{\infty}\frac{\bm{f}_s (-1)^s}{s!},
\end{equation}
where $\bm{f}_s$ is the $s$-th (vector) factorial moment of $\bm{P}(m)$,
\begin{equation}
    \label{fracmom-MMPPMI-app}
    \bm{f}_{s}=s!\sum_{m=s}^{\infty}\binom{m}{s}\bm{P}(m), \quad s=0,1,2,\dots.
\end{equation}
The series in Eqs. (\ref{P0-MMPPMI-app}) and (\ref{fracmom-MMPPMI-app}) both converge for any integer $s$, which was proved in Ref.~\cite{Ocinneide_1986}. From Eq. (\ref{fracmom-MMPPMI-app}),
\begin{equation}
    \bm{f}_0=\sum_{m=0}^{\infty}\bm{P}(m)=\bm{P},
\end{equation}
where $\bm{P}$ is the steady-state probability vector of the Markov process whose transition matrix is $D_0+D_1$. This means that $\bm{P}$ and therefore $\bm{f}_0$ can be computed by solving the steady-state master equation $\bm{P}(D_0+D_1)=0$. The other factorial moments can be computed by multiplying Eq. (\ref{Pm-MMPPMI}) by $m(m-1)\dots(m-s+1)$ and summing over $m$, which yields
\begin{equation}
    \label{fracmom-eq-MMPPMI-app}
    \bm{f}_s=s\bm{f}_{s-1}D_1(s\lambda I-D_0-D_1)^{-1},\quad s=1,2,3\dots.
\end{equation}
The stationary queue length distribution can be computed approximately provided enough factorial moments can be computed, see main text and Ref.~\cite{Ocinneide_1986} for further details. 

The moments of the stationary queue length distribution can be computed from Eq. (\ref{fracmom-MMPPMI-app}). The mean and the variance of the queue length are given by
\begin{subequations}
    \begin{align}
    & \mu=\bm{f}_1\bm{1}^T=\bm{P}D_1(\lambda I-D)^{-1}\bm{1}^T,\\
    & \sigma^{2}=\bm{f}_2\bm{1}^T+\mu-\mu^{2}=2\bm{P}D_1(\lambda I-D)^{-1}D_1(2\lambda I-D)^{-1}\bm{1}^{T}+\mu-\mu^{2},
\end{align}
\end{subequations}
where $D=D_0+D_1$. These expressions can be further simplified by means of the Neumann series,
\begin{equation}
    \sum_{n=0}^{\infty}A^n=(I-A)^{-1},
\end{equation}
which is valid for any square matrix $A$ such that  $\text{det}A<1$. Applying this result to $(xI-D)^{-1}$ for $x\neq 0$ and noting that $D\bm{1}^T=0$ and $\text{det}D=0$ (hence $\text{det}D/x<1$), we get
\begin{equation}
    D_1(xI-D)^{-1}\bm{1}^T=\frac{1}{x}\sum_{n=0}^{\infty}(-1)^n\frac{D_1 D^n\bm{1}^T}{x^n}=\frac{D_1\bm{1}^T}{x}.
\end{equation}
Using this identity, the mean and the variance of the queue length simplify to
\begin{subequations}
    \begin{align}
        & \mu=\frac{\bm{P}(D_1\bm{1}^T)}{\lambda},\\
        & \sigma^{2}=\frac{\bm{P}D_1(\lambda I-D)^{-1}(D_1\bm{1}^T)}{\lambda}+\mu-\mu^{2}.
\end{align}
\end{subequations}
The results for $\mu$ and $\sigma^2$ in this form were derived using the master equation approach in Ref.~\cite{Sanchez_2008} for a general stochastic gene expression model that is equivalent to the $MMPP/M/\infty$ queue discussed above.

\end{document}